\documentstyle[psfig,referee]{mn}
\newcommand{\beq}{\begin{equation}}
\newcommand{\eeq}{\end{equation}}
\newcommand{\bdm}{\begin{displaymath}}
\newcommand{\edm}{\end{displaymath}}
\newcommand{\bfig}{\begin{figure}}
\newcommand{\efig}{\end{figure}}
\newcommand{\msun}{M_{\odot}}
\def\etal{{\it et al.~}}
\def\ie{{\frenchspacing\it i.e. }}
\def\eg{{\frenchspacing\it e.g. }}

\def\gtsima{$\; \buildrel > \over \sim \;$}
\def\ltsima{$\; \buildrel < \over \sim \;$}
\def\prosima{$\; \buildrel \propto \over \sim \;$}
\def\gsim{\lower.5ex\hbox{\gtsima}}
\def\lsim{\lower.5ex\hbox{\ltsima}}
\def\simgt{\lower.5ex\hbox{\gtsima}}
\def\simlt{\lower.5ex\hbox{\ltsima}}
\def\simpr{\lower.5ex\hbox{\prosima}}

\def\HH{H$_2$~}
\def\HHP{H$_2^+$}
\def\gcc{g~cm$^{-3}$}
\def\percc{cm$^{-3}$}
\def\fhh{f_{\rm H_2}}
\def\tcentr{T_{\rm c}}
\def\ergs{\rm erg\;s^{-1}}

\title[Radiation from the first forming stars]{
Radiation from the first forming stars}

\author[E. Ripamonti, F. Haardt, A. Ferrara, \& M. Colpi]{E. Ripamonti$^1$, F.
Haardt$^2$, A.  Ferrara$^3$ \& M. Colpi$^4$\\
$^1$ Universit\`a degli Studi di Milano, Dipartimento di Fisica,
                    Via Celoria 16, 20133 Milano, Italy \\
$^2$ Universita' dell'Insubria, Dipartimento di
Scienze Chimiche Fisiche e Matematiche, Via Valleggio 11, 22100 Como, Italy\\
$^3$ Osservatorio Astrofisico di Arcetri, Largo Enrico Fermi 5,
      50125 Firenze, Italy \\ 
$^4$ Universita' degli Studi Milano-Bicocca, Dipartimento di
Fisica G.Occhialini, Piazza delle Scienze 3, 20126 Milano, Italy}
\date{June 2001}
\begin{document}
\maketitle

\begin{abstract}
The evolution of radiation emitted during the dynamical collapse of
metal-free protostellar clouds is investigated within a spherically
symmetric hydrodynamical scheme that includes the transfer of radiation
and the chemistry of the primordial gas. The cloud centre collapses on a
time scale of $\sim 10^{5-6}$ years, thanks to line cooling from 
molecular hydrogen (H$_2$). For most of the collapse time, when the
evolution proceeds self-similarly, the luminosity slowly rises up to
$\sim 10^{36}\;\ergs$ and is essentially due to \HH IR line
emission. Later, continuum IR radiation provides an additional
contribution, which is mostly due to the accretion of an infalling
envelope upon a small hydrostatic protostellar core which develops in the
centre. We follow the beginning of the accretion phase, when the
enormous accretion rate ($\sim 0.1 \;\msun\,{\rm yr}^{-1}$) produces a
very high continuum luminosity of $\sim10^{36}\;\ergs$. Despite the
high luminosities, the radiation field is unable to affect the gas
dynamics during the collapse and the first phases of
accretion, because the opacity of the infalling gas is too small; this
is quite different from present-day star formation. We also find that
the protostellar evolution is similar among clouds with different
initial configurations, including those resulting from 3D cosmological
simulations of primordial objects; in particular, the shape of the
molecular spectra is quite universal. Finally, we briefly discuss 
the detectability of this pristine cosmic star formation activity.

\end{abstract}

\section{Introduction}

As the temperature of the cosmic bath decreases, atoms start to
recombine and therefore decouple from the CMB radiation at redshift $\approx
1100$.  The baryonic Jeans mass immediately after this event is given by
\begin{equation}
\label{mj}
M_J \simeq 4.5\times 10^4 \left({\Omega_b\over \Omega}\right) (\Omega h^2)^{-1/2} \msun,
\end{equation}
where $\Omega$ and $\Omega_b$ are the
matter and baryon density parameters, respectively.  Masses larger than
$M_J$ are gravitationally unstable and should, in principle,
collapse. However, in order for the actual collapse to occur a more
severe condition must be satisfied, \ie that the cooling time of the gas
is shorter than the Hubble time at that epoch. In fact, radiative losses
provide the only way for the gas to lose pressure and settle down in the
potential well of the host dark matter halo.  Since the virial
temperature corresponding to the masses of the first objects (PopIII) is
typically $\simlt 8000$~K, cooling by hydrogen Ly$\alpha$ excitation is
strongly quenched, and the only viable coolant in a primordial H-He
plasma is molecular hydrogen.  \HH is produced during the recombination
phase, but its relic abundance is very small ($\fhh \approx
10^{-6}$).  This primordial fraction is usually lower than the one
required to fulfill the above time-scale constraint, but during the
collapse phase the molecular hydrogen abundance reaches values that are
high enough to allow the collapse to continue.  It follows that the fate
of a virialized lump depends crucially on its ability to rapidly
increase its \HH content during the collapse phase.

Haiman, Thoul \& Loeb (1996) and Tegmark \etal (1997) have addressed
this question by calculating the evolution of the \HH abundance for
different halo masses and initial conditions for a standard CDM
cosmology.  They conclude that if the prevailing conditions are such
that a molecular hydrogen fraction of order of $\fhh\approx 5 \times
10^{-4}$ can be produced during the collapse, then the lump will cool,
fragment and eventually form stars. This criterion is met only by the
largest PopIIIs, implying that at each virialization redshift one can
define a critical mass, $M_{crit}$, such that only protogalaxies with
total mass $M>M_{crit}$ will be able to eventually form stars.  As an
example, a 3$\sigma$ fluctuation of a Cold Dark Matter primordial
spectrum, has $M_{crit}\approx 10^6 \msun$ and collapses at $z\approx
30$.  Other studies have revised the value of $M_{crit}$ as a result of
a more accurate treatment of the micro-physics (Nishi \& Susa 1999;
Fuller \& Couchman 2000; Machacek, Bryan \& Abel 2001); however, these
results do not qualitatively alter the basic argument given above.

Recently, some works have started to tackle the formation and collapse
of PopIII objects through numerical simulations (Abel \etal 1998;
Nakamura \& Umemura 1999; Nakamura \& Umemura 2001; Bromm, Coppi \&
Larson 1999, 2001; Abel, Bryan \& Norman 2000 - ABN00; Bromm, Coppi \&
Larson 2001 - BCL01; Bromm \etal 2001) based on hierarchical scenarios
of structure formation.  These studies have shown that gravitational
collapse induces fragmentation of the first pregalactic objects with
initial baryonic mass $\approx 10^5 \msun$ into smaller clumps of
typical mass of about $10^{2-3} \msun$, which corresponds to the Jeans
mass set by molecular hydrogen cooling.

Tracking the subsequent evolution of these clumps is a very challenging
problem as it requires the simultaneous solution of the hydrodynamic
equations and of radiative transfer in lines. At present, due
to the tremendous computational demand, this as only been possible in
one dimensional formulations of the problem.  Star formation in the
early universe naively appears easier to understand as compared to the present, 
because several
complicating effects can be neglected to a first approximation: among
these are magnetic fields, dust grains and metal enrichment. 

In their pioneering work, Omukai \& Nishi (1998) (ON98) showed that at a
certain stage of the collapse a quasi-hydrostatic central core is formed
whose specific mass is almost independent on the details of the problem,
with a typical value of $\sim5\times10^{-3} \msun$. However, as the
accretion rate of the infalling gas is found to be rather high, around
$10^{-2} \msun$~yr$^{-1}$, the core mass at its formation might be
unimportant in setting the final mass of the star. In the absence
of any effect quenching accretion, a large fraction of the initial
object can become soon part of the protostar. Pushing this conclusion a
bit further, one might predict a top heavy IMF for this first generation
of sources.

If the infall can instead be stopped, different scenarios can be
envisaged.  It has yet to be firmly established if standard ways
proposed to halt the infall continue to work under primordial
conditions: radiation force might be opacity limited; bipolar flows need
some MHD acceleration process and therefore seem to be excluded by the
weak magnetic field present at the main formation epoch (Gnedin, Ferrara
\& Zweibel 2000), although the magnetic field could be amplified during
the protostar formation process. Angular momentum barrier related to
disk formation and competitive accretion among different clumps are
slightly more promising, but much more difficult to model
accurately. The estimate of the first mechanism (radiation reaction)
requires to accurately derive the properties of the radiation field
produced during the collapse and to evaluate the effects of the
radiation force on the infalling envelope. This is one of the main aims
of the present paper.

The second motivation for our study consists in assessing the detection
chances of distant PopIII objects via the (infrared) radiation they
release during the collapse phase. Ciardi \& Ferrara (2001)
have already pointed out that IR \HH molecular lines produced by the
cooling of supernova shocked PopIII gas will be very likely observed by
the {\it Next Generation Space Telescope}.  However, as \HH
roto-vibrational lines are essentially the only carrier of the entire
gravitational energy release, it seems worthwhile to compute the
intensity, shape and evolution of the emitted spectrum in detail.
Recently, Kamaya \& Silk (2001) (KS01) have estimated the \HH luminosity
from primordial collapsing stars using a very approximated treatment.

To investigate the described problem we have performed 1D numerical
simulations including a detailed treatment of the molecular line
radiative transfer.  The study is germane to the one of ON98 but it
includes the radiative force term in the dynamical equations adding a
number of imporvement, as we outline throughout the paper.

The paper is organized as follows.  In \S 2, we present our numerical
methodology.  The choice and justification for the initial conditions,
and the results of the simulation are discussed in \S 3. In \S 4 we
discuss the luminosities and the detectability, and in the final section
we give a brief summary.

\section{Numerical methods: the code}
We developed a 1-D Lagrangian hydrodynamical code, including full
treatment of radiative transfer and chemical evolution (with careful  
treatment of the \HH molecule, which is the main coolant in the
initial phases of the collapse). Such code is similar to the one
described by ON98. Here we only summarize the relevant differences
between the two; a more detailed description is given in the Appendix.

\subsection{Dynamics: radiative force}
We included the radiative force term into the momentum
equation, which then becomes

\begin{equation}
\label{momentum_one}
{{dv}\over{dt}} = -4 \pi r^2 {{dp}\over{dM_r}} - {{GM_r}\over{r^2}} +
f_{\rm rad} ,
\end{equation}
where we are exploiting the spherical symmetry, and where
\begin{equation}
\label{radforce}
f_{\rm rad}={1\over c}\int{\kappa_\nu F_\nu d\nu}
\end{equation}
is the radiative force per unit mass, $F_\nu$ is the specific energy
flux at frequency $\nu$ (both $f_{\rm rad}$ and $F_\nu$ are considered
in the radial outward direction), and $\kappa_\nu$ is the opacity at the
frequency $\nu$.

\subsection{Chemistry and thermodynamics: non ideal effects}
As discussed by several authors (\eg\ Hummer \& Mihalas, 1988; Mihalas,
D\"appen \& Hummer, 1988; Saumon, Chabrier \& Van Horn, 1995; hereafter
SCVH95), at high densities ($\rho\gsim10^{-3}\;{\rm
g\,cm^{-3}}$) particles are jammed closely together, and the bound
electron orbitals filling too large a volume fail to survive so that
electrons migrate from atom to atom. Such an effect is known as
\emph{pressure ionization} and generally it is not taken into account in
the Saha equations.  As an example, a pure hydrogen gas with $\rho=1$ g
cm$^{-3}$ and $T=3\times10^4$ K is largely ionized (the mean
distance between nuclei is only twice the Bohr radius of an H atom on
the fundamental level, and the electrons are free in a Fermi gas,
interacting simultaneously with several ions), while solving the Saha
equations results in a completely molecular gas, which is clearly
unphysical. In such regimes, the gas is also non-ideal, and the perfect
gas equations are poor approximations.

For these reasons, at high densities and temperatures we find the
chemical abundances and thermodynamical properties (such as the pressure
$p$, the internal energy per unit mass $u$ and the entropy per unit mass
$s$) by interpolation of the values tabulated by SCVH95, which are
calculated using the \emph{free energy minimization} technique.  SCVH95
calculations include several non-ideal effects, such as, (a) the partial
electron degeneracy, (b) the Coulomb interactions among charged
particles, (c) the finite extension of ``hard-sphere'' particles, (d)
weak attractive interactions (\eg van der Waals), and (e) modifications
of the internal partition functions due to the shift of the energy
levels.

\begin{figure*}
\psfig{figure=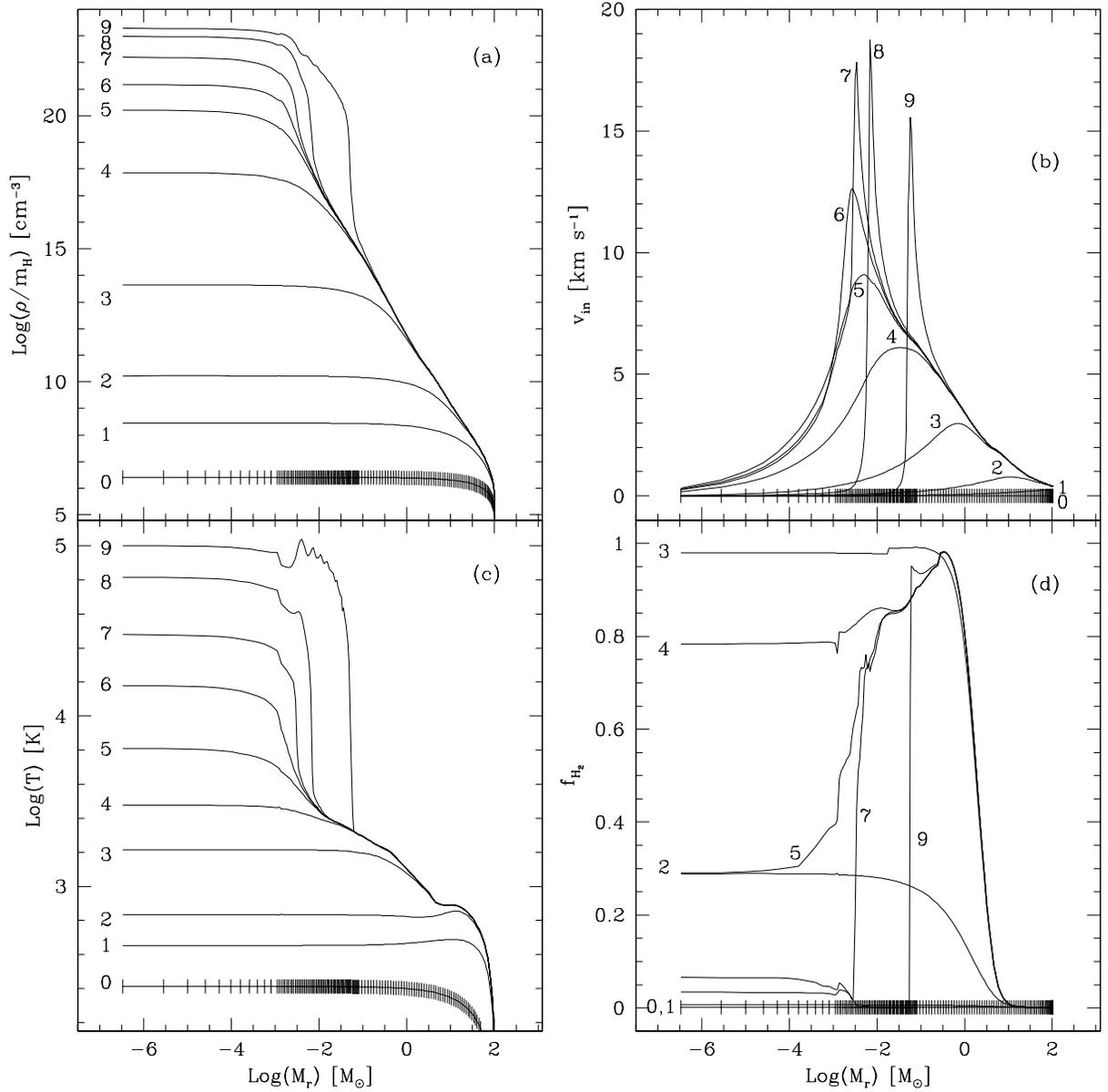,height=17.8cm,width=17.8cm}
\caption{Spatial structure of the cloud in run P100/2 at 10 different
stages of collapse (see Table \ref{collapse_st}). Panels refer to
density (a), infall velocity (b), temperature (c) and to the fraction of H
nuclei inside \HH molecules (d). On the abscissa we use the mass
enclosed within radius r ($M_r$). Shell edges are denoted by small
tags onto the curve referring to the initial configuration; in panel (d)
we omit stages 6 and 8 for clarity. The small oscillations in the
temperature profile of stage 9 around $M_r\simeq-2$ are due to numerical
effects.}
\label{ss1}
\end{figure*}


\section{The simulations}

\subsection{Initial conditions}

The main parameters specifying the initial condition of our runs are: the
total cloud mass $M_{\rm tot}$, the temperature, density, and velocity
profiles $T_0(M_r)$, $\rho_0(M_r)$, $v_0(M_r)$, respectively, and the
hydrogen molecular fraction $\fhh$.

Several previous investigations set constraints on such parameters.
Theoretical predictions of the mass of fragments leading to the birth of
the first stars are quite uncertain, ranging from a minimum of $\sim
0.1\; \msun$ (Palla, Salpeter \& Stahler 1983), to $\gsim 200\; \msun$
(Hutchins 1976).  On the other hand, there is agreement about the value
of $\fhh \sim 5 \times 10^{-4}$, and about the value of $T_0(M_r)$ which
is of order of few hundreds degrees (\eg Tegmark \etal 1997).  Detailed
numerical simulations (BCL01, ABN00) suggest that the runaway collapse
starts soon after the (number) density reaches a value of $10^4-10^5$
\percc. Such value is determined by the transition from non-LTE cooling
(with $\Lambda\propto n^2$) to LTE cooling (with $\Lambda\propto n$),
slowing down the contraction, and allowing the gas to reach a
quasi-hydrostatic state.  This transition occurs when $T\sim 300-400$~K
and $\fhh\sim5-10\times 10^{-4}$, inside a cloud with a mass of a few
thousand $\msun$, a value close to the Jeans mass of the cloud.

To check how our results depend on the choice of the initial conditions,
we run a set of models with different initial mass $M_{\rm tot}$,
ranging from $10$ to $\sim1500\;\msun$. We use a constant profile for
the \HH abundance and zero velocity (the cloud is at rest).  The density
and temperature profiles inside the cloud are derived imposing
hydrostatic equilibrium and a polytropic index equal to 5/3 or 1
(isothermal).  In addition we have considered `numerical' profiles as
given by ABN00 from cosmological simulations.  In runs P10, due to the
high value of the initial central density, we used a higher value for
$\fhh$. The initial electron fraction adopted is $f_{e^-}=10^{-10}$.
With the exception of run P100/2, we always divide our simulated cloud
into 100 Lagrangian shells, with massive external shells and much
lighter internal ones (\eg, in run P100 each external shell comprises
$\gsim1\;\msun$ of gas, while the central one has a mass of only
$3\times10^{-7}\;\msun$).  Such choice is aimed to follow with good
resolution the fast evolution of the centre of the cloud, which accounts
only for a tiny fraction of the total mass.  Details of the initial
conditions and general set up are given in Table \ref{initial_c}.  Note
that runs P100 and P10 are directly comparable to those of ON98. Run
P100/2 is identical to run P100, but the object is divided into 140
(rather than 100) shells; the ``extra'' shells were used to improve the
resolution in the particularly important region where the shock forms
($-3\lsim\log{M_r}\lsim-1$).

\begin{table}
\caption{Initial conditions}
\label{initial_c}
\begin{tabular}{cccccl}
 Run & $\rho_{\rm 0,c}$ & $T_{\rm 0,c}$ &
 $f_{\rm H_2}$ &Notes\\
     & ($m_H\;$cm$^{-3}$) & (K) & ($10^{-4}$)\\
\hline
 P100  & $2.6\times10^6$ & $270$ & $5$ &cfr. ON98(A,B)\\
 P100/2  & $2.6\times10^6$ & $270$ & $5$ &refinement of P100 (140 shells)\\
 P1000  & $2.3\times10^4$ & $246$ & $5$  &\\
 P10  & $3.5\times10^8$ & $249$ & $50$ &cfr. ON98(C)\\
 I100  & $7.8\times10^7$ & $300$ & $5$  &\\
 I1000  & $1.1\times10^7$ & $300$ & $5$  &\\
 N1500 & $2.0\times10^6$ & $440$ & $5$ &from ABN00\\
\hline
\end{tabular}
\medskip

The letter in the label of each run gives the assumed initial
temperature/density profile: polytropic, isothermal (I), and numerical
(N), \ie after ABN00 results. It is followed by a number which gives the
assumed total cloud mass $M_{\rm tot}$, in solar masses.
\end{table}

\begin{figure}
\psfig{figure=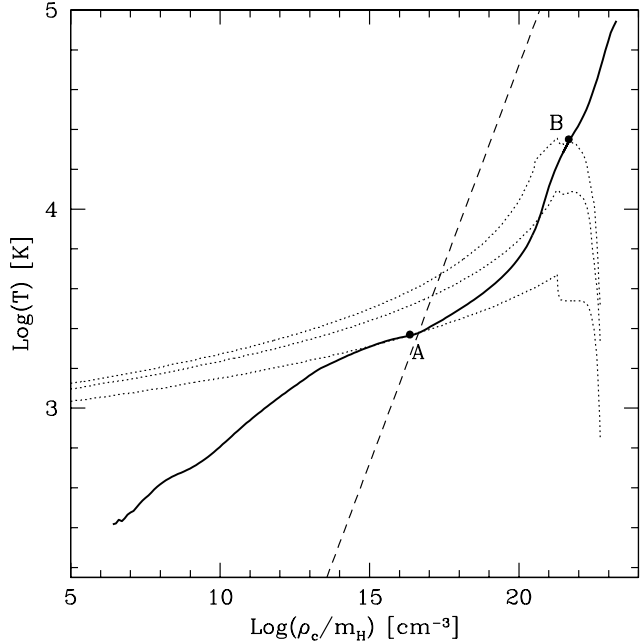,height=9cm}
\caption{Evolution of density and temperature of the innermost
shell (thick line). Points A and B mark conditions when the cloud
becomes optically thick and when the hydrostatic core forms,
respectively.
Thin dotted lines show the loci of points where
equilibrium \HH fraction is 0.9 (bottom), 0.1 (middle) and 0.01
(top). Dashed line show the theoretical locus of points where a
generical cloud should become optically thick.}
\label{t_rho}
\end{figure}

\subsection{Dynamical evolution}
The dynamics of the collapse is illustrated in Figure \ref{ss1} that
shows the evolution of the cloud in run P100/2 at 10 different stages
(see Table \ref{collapse_st}). In Figures \ref{t_rho} and \ref{chem_rho}
we show the evolution of the temperature and of the main hydrogen
species in the most central shell as a function of density. In the rest
of this section we will describe the results obtained in run P100/2;
results of other runs are quite similar.  In runs P100, P100/2 and P10
we find good agreement with the results of ON98.

In the beginning, the evolution is very well described by a
Larson-Penston self-similar solution (Larson 1969 - L69; Penston 1969)
with a density profile scaling as 
$\rho\propto r^{-\alpha}$ (with
$\alpha\simeq2.2$).
The self-similar evolution breaks down between stages 6 and 7, when a 
quasi-hydrostatic core forms in the centre and starts growing in mass.

\begin{table}
\caption{Collapse stages (run P100)}
\label{collapse_st}
\begin{tabular}{cccccl}
 Stage & $T_c$ & $\Delta t$ & $(t_f-t_i)$ &Notes\\
       & (K)     & (yr)          & (yr)           & \\
\hline
 0 &  270   & ---                 &  $740000$    &
Initial conditions \\
 1 &  450   & $720000$ 	&  $20000$    &
Start Rad. Trans.\\
 2 &  650   & $19000$ 	&  $930$    &
\\
 3 &  1500  & $920$ 	&  $9.6$    & \\
 4 &  3000  & $9.2$ 	&  $0.40$ & \HH is dissociating\\
 5 &  6500  & $0.060$   &  $0.34$ &\\
 6 &  15000 & $0.005$   &  $0.34$ & Core is forming
\\
 7 &  30000 & $0.002$   &  $0.34$ &
Core has formed\\
 8 &  65000 & $0.011$   &  $0.33$ &\\
 9 &  100000 & $0.33$   &  ---                 &
Final stage \\
\hline
\end{tabular}
\end{table}

\subsubsection{The collapse phase}
\begin{figure}
\psfig{figure=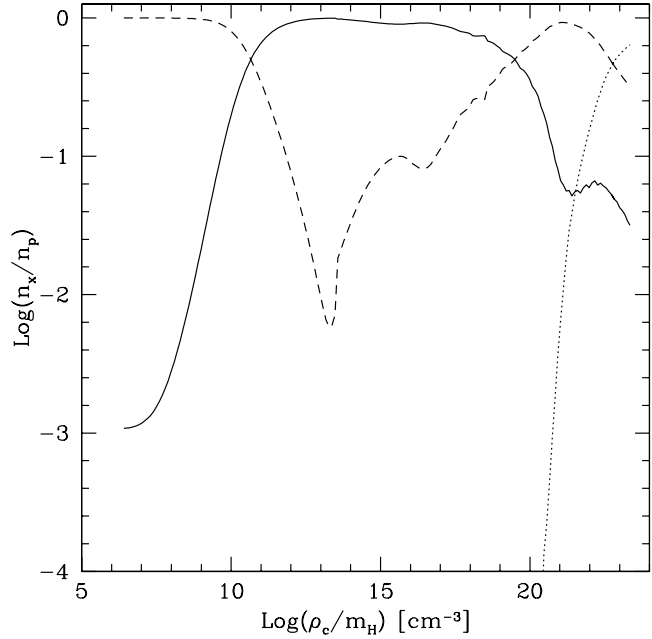,height=9cm}
\caption{Evolution of the fractions of the main hydrogen species in the
innermost shell as a function of density: \HH (solid line), H$^0$
(dashed line) and H$^+$ (dotted line).}
\label{chem_rho}
\end{figure}

\begin{figure}
\psfig{figure=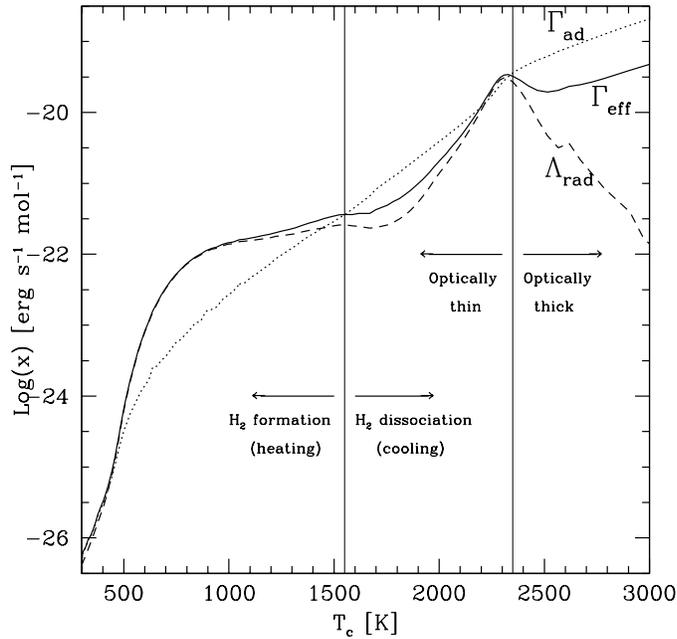,height=9.0cm}
\caption{Radiative cooling rate (dashed), adiabatic heating rate
(dotted) and effective (i.e. adiabatic plus chemical) heating rate
(solid) in the innermost shell as a function of its temperature. The
left vertical line marks the transition from H$_2$ formation to H$_2$
dissociation, while the right vertical line separates the optically thin
and optically thick regimes.}
\label{lambda_t}
\end{figure}

During collapse the central regions of the cloud undergo several
important chemical and thermal transformations.  When the number density
reaches $\gsim10^8$ \percc (stage 1), 3-body reactions become efficient,
converting the bulk of H into \HH (Fig. \ref{ss1}, stages 2-3;
Fig. \ref{chem_rho}). The growing \HH fraction increases the gas cooling
rate, only partially counterbalanced by radiative transfer effects (see
next section 3.3).  The evolution of the thermal properties of the gas
is shown in Figure \ref{lambda_t}, where the adiabatic heating
$\Gamma_{\rm ad}$, the radiative cooling rate $\Lambda_{\rm rad}$ and
the effective heating $\Gamma_{\rm eff}=\Gamma_{\rm ad}+\Gamma_{\rm ch}$
($\Gamma_{\rm ch}$ is the heating - or cooling - due to chemical
reactions) are plotted vs. the gas temperature. All these quantities
refer to the central shell.  As $T\lsim 1550$ K, the chemical heating
due to the \HH formation dominates over $\Gamma_{\rm ad}$ and heating
and cooling terms almost balance, leading to a very slow temperature
growth.  For $T \gsim 1550$ K \HH is destroyed rather than formed, and
the chemical term becomes a net cooling (this is the reason why the
curves $\Gamma_{\rm eff}$ and $\Gamma_{\rm ad}$ cross at $T\simeq 1550$
K).  At this stage the chemical and radiative cooling terms are of the
same order.  Radiative cooling is dominated by line radiation for
$T\lsim 1800$ K, while continuum emission is important above.  However
continuum radiative cooling soon becomes highly inefficient because of
the large optical depth: for $T\gsim 2350$ K the innermost regions
become almost adiabatic, and compressional heating is only matched by
chemical cooling.

We just mentioned that \HH starts
to dissociate when $T\gsim 1550$ K. This process is
slow because of the density increase and the chemical cooling associated
to the dissociation process (4.48 eV per dissociated molecule are lost
by the gas thermal energy).
As a result, the \HH fraction at the end of the self-similar phase 
($\tcentr\sim15000$ K, stage 6) is still $\simeq 5\%$, even in the
hottest central regions.

\subsubsection{Core formation}

During the last stages of the self-similar collapse (stages 5-6;
evolution from point A to point B in Fig. \ref{t_rho}), when the centre
of the cloud is essentially adiabatic, \HH dissociation acts as a
``thermostate'', preventing a faster increase in temperature and
pressure, which would halt the collapse.  In fact, dissociation absorbs
$\gsim 50$\% of the heat generated by gas compression, which (due to the
opaqueness of the cloud's centre) could not have been dissipated
otherwise.  When nearly all \HH molecules are
dissociated, the thermostatic effect is no longer effective, and the
central temperature and pressure rise rapidly stopping the collapse of
the central regions.  At the end of \HH dissociation ($\tcentr\sim20000$
K) the chemical energy and the thermal energy per H$^0$ atom are similar
($\simeq 2.2$ eV and $\simeq 2.6$ eV, respectively).  This fact reflects
the virial theorem: half of the cloud gravitational energy is converted
into thermal energy after compression, while the rest is tapped into
chemical energy through dissociation.






Outside the core, the gas is still collapsing with a velocity close to
free-fall. This results in a shock front forming at the edge of the
core, where the infalling material is suddenly halted (see
Fig. \ref{ss1}.b).  Unfortunately, the numerical method (\ie the
introduction of an artificial viscosity) used for the treatment of
hydrodynamical effects is inaccurate when dealing with strong shocks,
since they are spread upon several shells. However, though the treatment
of the shock physics is only approximate, the general picture of the
cloud evolution is essentially correct. In run P100/2 we also improve
the situation by refining the shells in the region where the shock forms.

The core evolution is driven by two main processes.
First, as can be seen in Figures \ref{ss1}.b, and
\ref{core_mass}, the core mass (defined as the mass inside the radius
where the infall velocity is $<0.1v_{\rm ff}$, $v_{\rm
ff}\equiv\sqrt{2GM_r/r}$) grows, due to the
accretion of infalling material.  Second, even if the infall velocity is
drastically reduced, the core is compressed by the infalling material,
and then heats up (as can be seen in Fig. \ref{t_rho}, beyond point B).

\begin{figure}
\psfig{figure=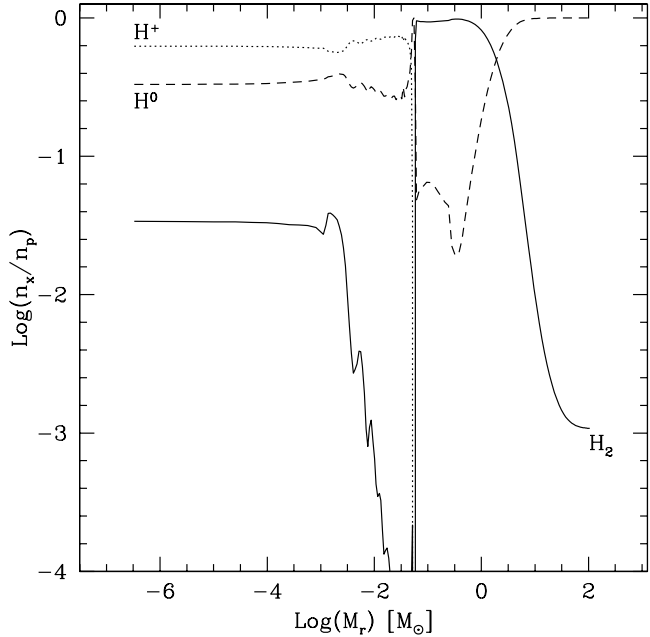,height=9.cm}
\caption{Chemical structure of the cloud at stage 9: \HH (solid line),
H$^0$ (dashed line) and H$^+$ (dot-dashed line); the ionized hydrostatic
core, the inner molecular and the outer atomic envelope are quite
apparent; the peak of H$^0$ curve at $\log{M_r}\sim-1.3$ is located
in the region where the (smoothed) shock brings down the infall velocity
of the shell from the maximum value of $\sim 15\;{\rm km\,s^{-1}}$ to
$\sim 2\;{\rm km\,s^{-1}}$. The small oscillations for $-2.5\lsim
\log{M_r}\lsim-1.5$ are numerical.}
\label{chem_str}
\end{figure}

\subsubsection{Cloud structure after core formation}

After core formation and during its subsequent evolution, the \HH
fraction in the centre never rises again to important
levels\footnote{The central value of $\fhh$ at late stages is quite
uncertain: using SCVH95 results we obtain $\fhh\sim0.03$; had we used
results such as those of Mihalas, D\"appen \& Hummer 1988, we would have
obtained $\fhh < 0.01$; ON98 find $\fhh\sim0.4$, but this is likely to
be too high since they do not consider pressure ionization effects.}.
\HH tends to form a layer of several solar masses between the small core
(where H is progressively ionized) and the outer layers (where H is
essentially atomic, see Figure \ref{chem_str}).  After stage 6 we can
recognize three different regions, i.e., the core, the molecular
envelope, and the outer region
(see Figure \ref{chem_str}).
 
\begin{itemize}
\item[(1)]{The central core is a region whose infall velocity is much
smaller than the free-fall value. At core formation, the gas here is
basically neutral, but at stage 9 it has become mostly ionized (but
$n_{\rm H^+}/n_{\rm H}$ is still only $\sim 2$) because of the pressure
ionization process.  The outer edge of the core is marked by a shock
front, where material accretes nearly with free-fall velocity.  Since
the artificial viscosity technique ``smooths'' the shock, we have a thin
transition region, where neutral H is the dominant species; that may be
an numerical artefact.}

\item[(2)]{Outside the shock defining the core outer edge, there is an
envelope of $\sim3\;\msun$ composed by molecular hydrogen
(the shock also marks a sharp transition in chemical composition,
from \HH to H$^0$ or H$^{+}$). 
Such envelope is essentially in free fall ($v/v_{\rm ff}$ is in the
range 0.3-0.7, decreasing outward). Its outer edge is roughly at a
radius $\gsim10^{15}$ cm.}

\item[(3)]{The remaining outer gas, accounting for most of the cloud mass, 
has a chemical composition only slightly different from the initial
one, since reactions leading to \HH formation are slow 
(mainly because of the low density), and involve only a minor fraction 
of H atoms.
This layer is  the  extension to the inner molecular envelope, 
and has a velocity  $v/v_{\rm ff}$ between 0.1 and 0.3.}
\end{itemize}

At stage 9 the core temperature and density are $\simeq 10^5$ K and
$\simeq 0.3$ \gcc, respectively. In these conditions the Courant
time-scale is extremely short ($\lsim 300$ sec.), and it becomes more and
more difficult (and costly, in terms of CPU time) to follow the
evolution further.  For this reason, we stopped our calculations when the
central temperature reached $\sim10^5$ K.

\subsubsection{The accretion phase}

After core formation, the evolution continues because of
the accretion of the outer envelope onto the core.

Despite our  approximate treatment of the shock,
we find that the accretion rate, as inferred from the full calculation, is extremely high:
the core grows from $\sim0.3\times10^{-2}\;\msun$ when it forms, to
$\sim4.5\times10^{-2}\;\msun$ in $\sim 0.30$ yr, which means
$\dot{M}\sim0.14\;\msun$ yr$^{-1}$.  Such accretion rate is much larger
than typical estimates for present-day star formation
($\dot{M}\lsim10^{-4}\;\msun$ yr$^{-1}$).

In order to extend our results to a more advanced state of accretion, we
modified our code  ``freezing'' the evolution of the shells deeply
embedded in the core. In practice, when a shell has i) a small infall
velocity ($v/v_{\rm ff}<10^{-3}$), ii) a high temperature
($>5\times10^4$ K), and iii) is located at least 5 shells inside the
shell where the infall velocity is maximum, we completely stop its
evolution ({\frenchspacing\it i.e.}, we set velocity at 0 and keep all
the other quantities constant to the last computed values).

This greatly speeds-up our calculations because there is no more need to
follow the shells with the shortest time-scales,
drastically reducing the number of integration steps needed. Such
``freezing'' is appropriate, since these 
shells do evolve on time-scales much longer than those
imposed  by the Courant condition and have little
influence upon the layers where  accretion occurs.
There are two major concerns:
\begin{itemize}
\item[-]{the neglect of the secular changes in the core radius: for
example, in the full calculation the radius of the 10 innermost shells
undergoes a reduction of about $40$\% (from $1.6\times10^{10}$ cm to
$1.0\times10^{10}$ cm) as $T_{\rm c}$ goes from $50000$ K (stage 8) to
$\simeq10^5$ K (stage 9), while in the simplified calculation performed
between stages 8 and 9, we keep this radius constant. This
change is less important in more external shells: the difference in
the shock position between the ``full'' and the ``frozen'' run at stage
9 is found to be $\simeq$4\%.}
\item[-]{the radiation produced upon the shock surface could ionize a
thin shell outside the shock itself, as reported in Staher, Palla \&
Salpeter (1986) (SPS86), modifying the dynamics of accretion (see next
sections).}
\end{itemize}

\begin{figure}
\psfig{figure=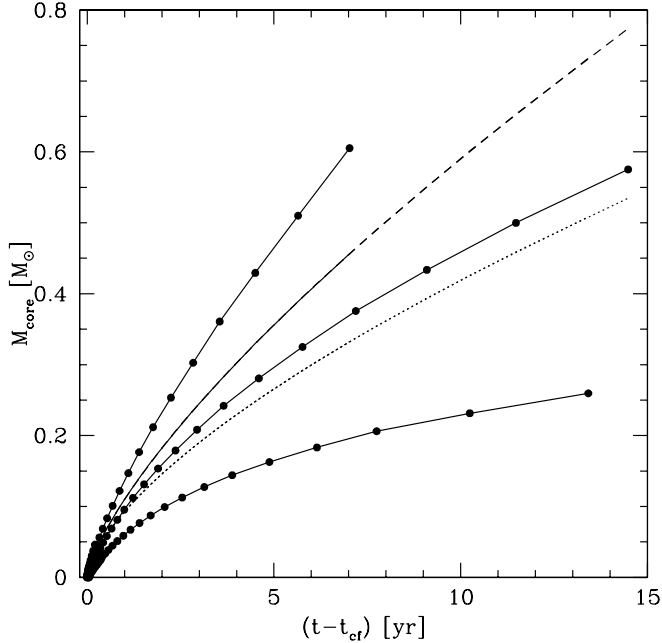,height=9.cm}
\caption{Evolution of the core mass. The three lines with dots
refer to model P1000 (top line), P100 (middle line) and P10 (bottom
line). The dashed line shows
the predictions of ON98 (eq. 11), while the dotted one
shows the results of the calculation described in the text, using the
parameters of run P100.}
\label{core_mass}
\end{figure}

Instead, the influence of the lowered temperatures in central
shells is completely negligible, since the optical depth from the
``frozen'' shells is very high ($\tau_{\rm c}\gsim 10^{12}$).

Figure \ref{core_mass} shows the result of the accretion calculations
performed with the modified code: we stop $\gsim10$ years after core
formation. In the first 0.34 years this results are almost identical to
those obtained with the ``complete'' code; instead, points at later
times should be considered indicative.

ON98 investigated the evolution of the mass accretion rate on the newly
formed hydrostatic core using the Larson-Penston self-similar solution
(L69, Penston 1969) and its extensions after core formation (Hunter
1977, Yahil 1983, Suto \& Silk 1988).  For a model close to our model
P100 (and P100/2), they found that the core mass grows following a
power-law
\begin{equation}
\label{mcore_on11}
M_*(t)=0.11\;\msun ((t-t_{\rm cf})/{\rm {yr}})^{0.73}.
\end{equation}

We also find an empirical model which could better fit the numerical
results, based on the fact that in our calculations we find that the
mass flux $M_{\rm flux}(M_r)=4\pi r^2 \rho v_{\rm in}$ is well
approximated by a power-law. Such empirical model leads us to predict that

\begin{equation}
\label{mcore_diffsol}
M_*(t)=9.2\times10^{-2}\msun\left({{t\over{\rm {yr}}}+0.0063}\right)^{0.657}.
\end{equation}
Both these analytical expressions are compared with the numerical results
in Figure \ref{core_mass}.

\begin{figure}
\psfig{figure=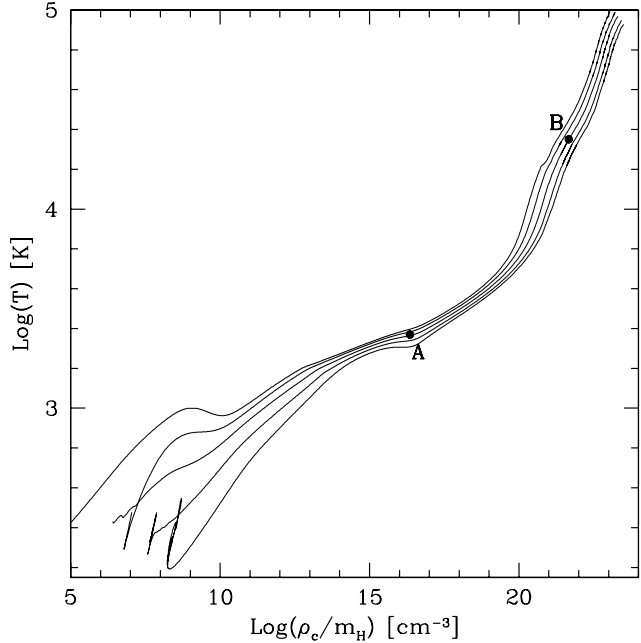,height=9.cm}
\caption{Comparison of the evolution of central conditions (density and
temperature) in the several of our computed models: from top to bottom,
the lines refer to model P1000, I1000, P100, I100 and P10; model N1500
is almost indistinguishable from model P100, except in the first
stages. As in Fig. \ref{core_mass}, points A and B denote the conditions
where (in model P100) $\tau$ first becomes $>1$ and where core formation
occurs, respectively.}
\label{evolutionary_tracks}
\end{figure}

\subsubsection{Importance of initial conditions}

The above description essentially applies to all
the models we computed: the evolutionary sequence is always the same,
even if there are small quantitative differences. Models lose memory of
the initial conditions almost completely during the self-similar phase,
as can be seen in Fig. \ref{evolutionary_tracks}, where tracks referring
to the different models are very close for $n_{\rm c}>10^{14}$ \percc, \ie
well before the cloud becomes optically thick (point A) or forms the
central hydrostatic core (point B).

There remain only relatively small differences (within a factor of 2 or
3), in particular:
\begin{itemize}
\item[-]{Models with higher total mass ($M_{\rm tot}$), have higher
temperatures than low $M_{\rm tot}$ models, for any given central
density $n_{\rm c}$;}
\item[-]{Models with high $M_{\rm tot}$  have cores that grow
faster than low
$M_{\rm tot}$ models (see Fig. \ref{core_mass}).}
\end{itemize}

Both these trends can be easily explained: when we impose that in
the initial stage the clouds are close to equilibrium, we implicitly
force high-mass models to have higher temperatures at any given density;
even if quantitative differences are strongly reduced in the
self-similar phase, high-mass models remain warmer than low-mass models
at all stages. Furthermore, the accretion rate depends on temperature,
since $\dot{M}\sim M_{\rm Jeans}(T,\rho)/t_{\rm
ff}(\rho)\propto T^{3/2}$ (see Stahler, Shu \& Taam, 1980) and
temperature differences explain also those in $\dot{M}$.

The behaviour of the N1500 model (the only one for which we do not impose
quasi-equilibrium initial conditions) is particularly relevant:
after some ``wandering'' in the $n_{\rm c}-T_{\rm c}$ plane,
its central conditions evolve almost exactly along the evolutionary
track of model P100, indicating that the collapse does not preserve the
memory of initial conditions and the cloud evolves along an equilibrium
track.

\subsection{The radiation field}

The evolution of the radiation field is intimately coupled with dynamics
and our physical description of forming primordial stars fully accounts
for their emission properties.  Here we give details on the evolution of
the optical depth, luminosity and spectra during the two phases of
collapse and core formation, and at the very beginning of the accretion
phase.

\begin{figure*}
\hbox{
\psfig{figure=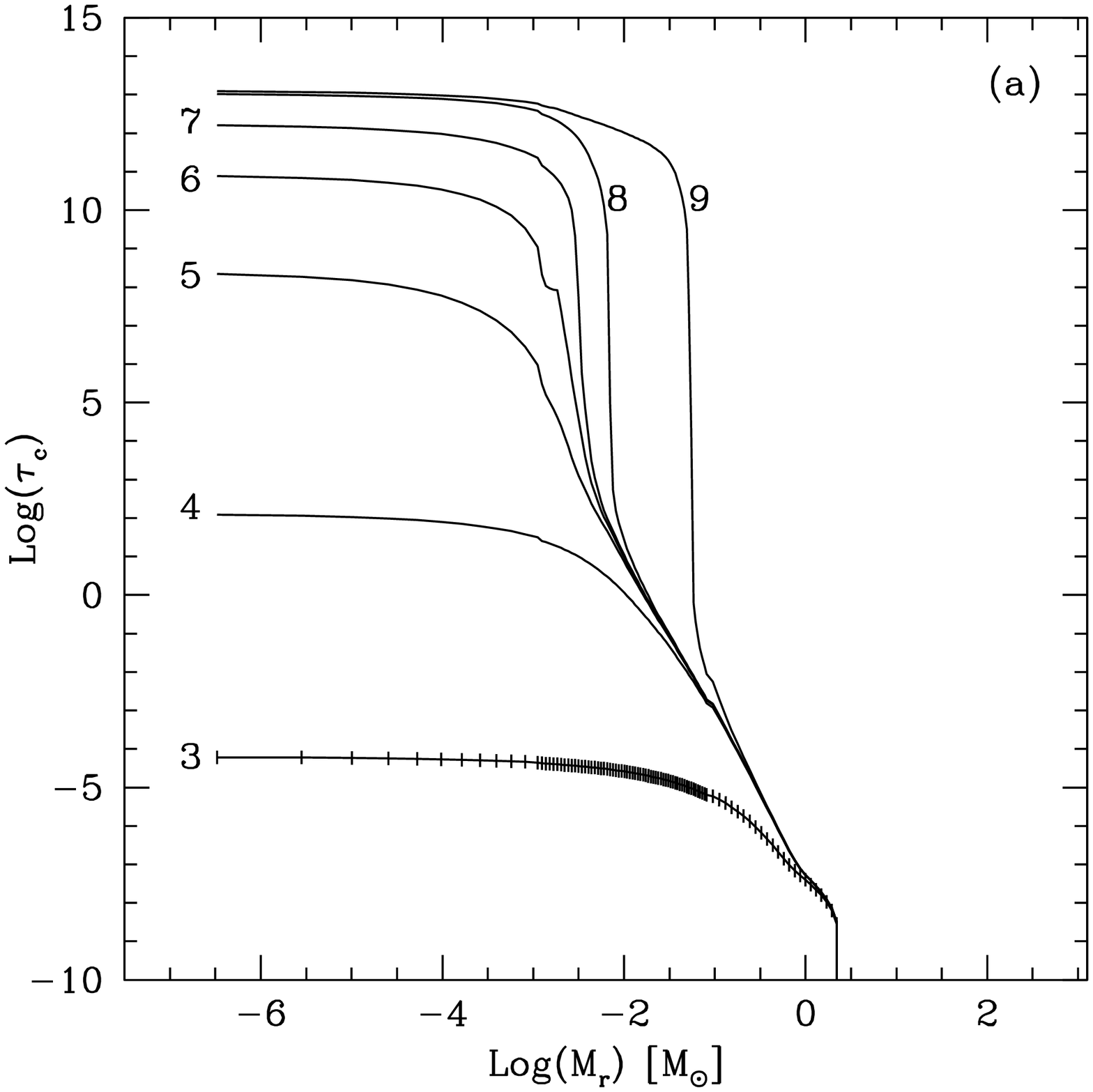,height=9.0cm}
\psfig{figure=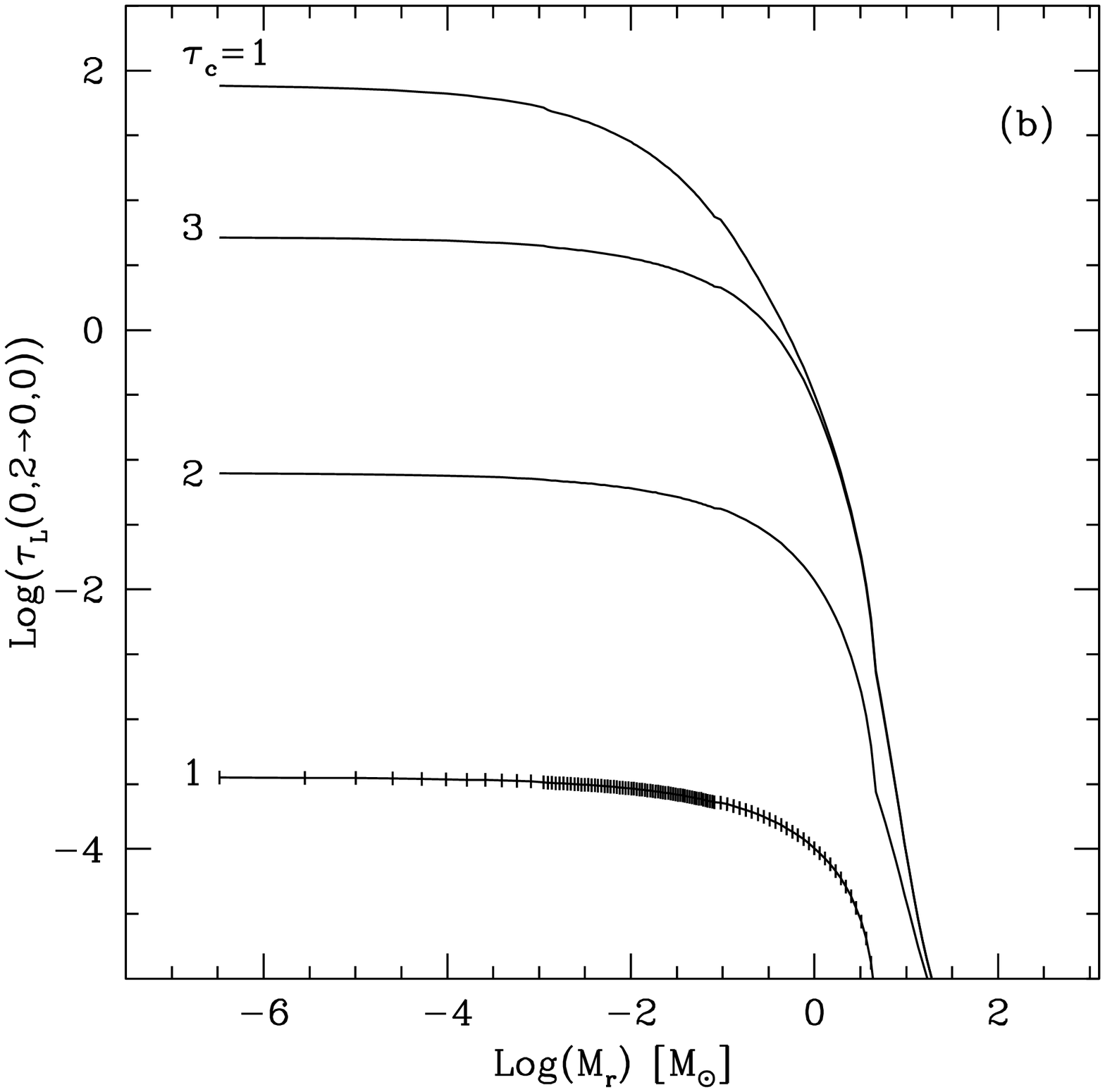,height=9.0cm}
}
\caption{Radial outward optical depth as a function of enclosed mass for
(a) continuum radiation and (b) for radiation at the center of the
roto-vibrational \HH (0,2)$\rightarrow$(0,0) transition
($\lambda=28.3\,\mu{\rm m}$); the top curve in this panel, taken when
$\tau_{\rm c}=1$, is represents all stages between 4 and 9.}
\label{tau_m}
\end{figure*}

\subsubsection{Luminosity from the collapsing cloud}

At the onset of dynamical evolution, radiation results from the cooling
of the optically thin \HH present in the cloud.  At stage 1, as
3-body reactions keep converting H atoms into \HH molecules, the line
optical depth rapidly grows, reaching significant values. At stage 2
there are already 14 lines with $\tau>0.1$ (although none of them
with $\tau=1$), and this number rises to 42 lines (half with
$\tau>1$) by the time the central temperature reaches $1000$ K
(between stage 2 and stage 3). In Figure \ref{tau_m}a we show, as an 
illustrative example, the evolution of the optical depth profile for 
a selected line during the self-similar phase.

The increase of the optical depth obviously
affects the gas cooling properties (as can be seen from the flattening
of the radiative cooling curve in Fig. \ref{lambda_t}), but it does not
influence the dynamics in a fundamental way. 
The increase in \HH abundance is
associated with an increase of the gas temperature and of the number 
of relevant transitions of the \HH molecule. 
As the temperature grows, new, high quantum number lines are initially 
optically thin, providing a minimum level of cooling. 
Though line cooling is severely quenched by opacity 
(e.g., at $\tcentr\simeq 2000$ K only a 
fraction $\simeq 0.02$ of the line flux is not re-absorbed locally)
its absolute value 
\emph{increases} anyhow as $\tcentr$ increases; 
this prevents the
temperature from rising too fast, possibly stopping the collapse in its early
phases.

As already mentioned, when $\tcentr=1000$ K we start to include cooling
from the continuum, which rapidly gains importance, becoming the
dominant process in the central $\sim0.1\;\msun$ already when
$\tcentr\sim1800$ K. The evolution of the continuum opacity profile is 
shown in Figure \ref{tau_m}b, where the formation of 
the core sharp edge is apparent. 

\begin{figure}
\psfig{figure=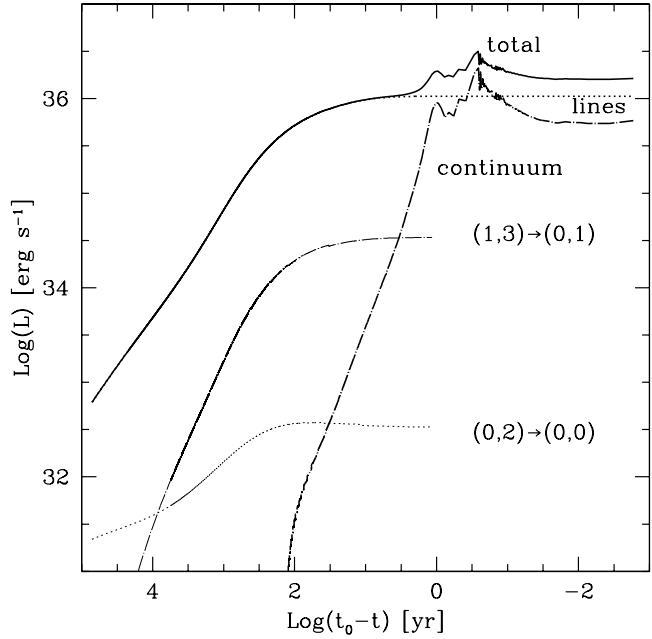,height=9.cm}
\caption{Evolution of the luminosity in lines (heavy dotted), continuum
(heavy dot-dashed) and their sum (thick solid line); thin lines refer to
two \HH transition: (0,2)$\rightarrow$(0,0), $\lambda=28.3\,\mu{\rm m}$
(dotted) and (1,3)$\rightarrow$(0,1), $\lambda=2.12\,\mu{\rm m}$
(dot-dashed); they do not appear when we ``freeze'' all line
luminosities. Note that the x-axis shows the time left before the end of
our calculation. The oscillations in the continuum luminosity between
$0.5\gsim \log(t_0-t)\gsim-1$ are connected with core formation and
could be partially numerical.}
\label{l_evol}
\end{figure}

\begin{figure}
\psfig{figure=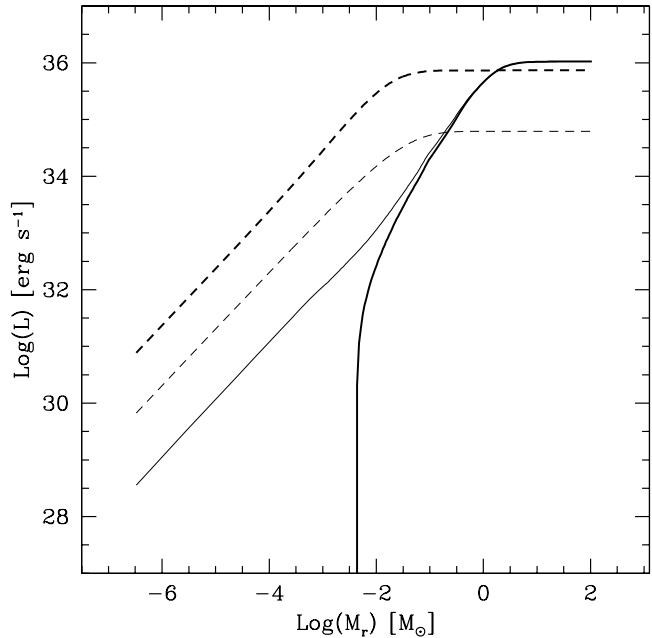,height=9.0cm}
\caption{Outward total luminosity in \HH lines (solid) and in the continuum
(dashed) as a function of enclosed mass at two different times: when
$\tcentr=2000$ K, $\tau_c\lsim0.1$ (thin lines) and when $\tcentr\sim2300$ K,
$\tau_c\sim10$ (thick lines).}
\label{lstruct}
\end{figure}

The corresponding emitted luminosities vs. time are plotted in Figure
\ref{l_evol}.  Note that even when continuum radiation dominates the
central emission, the \HH lines still have a comparable or slightly
larger overall luminosity. This is due to the fact that line radiation
comes from the outer, massive \HH envelope, while continuum comes from
an inner, much smaller region (only slightly larger than the core
itself); this can be seen in fig. \ref{lstruct}, where we show the
radial dependence of the emitted luminosity; in particular, note that
the drop in the central line luminosity has negligible influence upon
the total line luminosity, and that the growth of continuum luminosity
stops at $M_r\sim10^{-2}-10^{-1}\;\msun$, while line luminosity is still
growing at $M_r\sim1\;\msun$.

\subsubsection {Spectra}
\begin{figure}
\psfig{figure=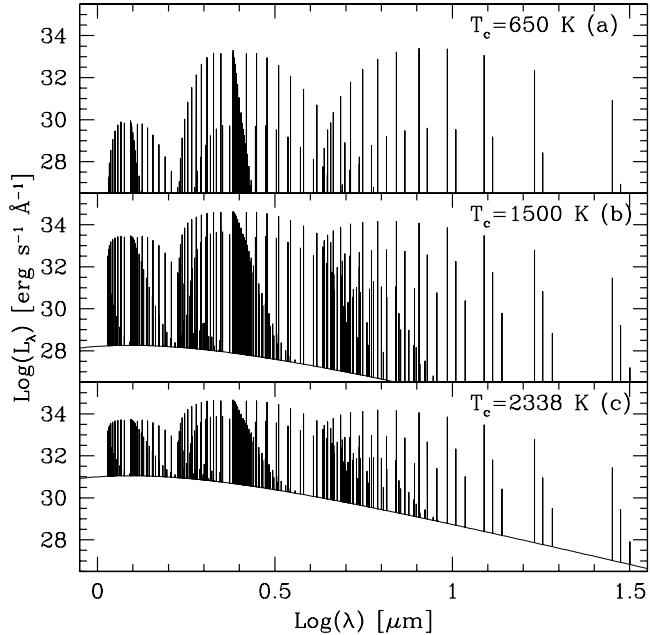,height=9.cm}
\caption{Emitted spectra when $T_{\rm c}$ is $650\;K$ (a), $1500\;K$ (b) and
$2338\;K$ ({\it i.e.} just before we ``freeze'' lines) (c). Line widths are
actually smaller ($\Delta\lambda/\lambda\lsim10^{-4}$) than line
thickness in this plot, so that overlaps are only apparent.}
\label{spectra}
\end{figure}

In Fig. \ref{spectra}, we show the emitted spectra at 3 different
stages.  In the upper panel the spectrum comes from line emission only,
while in the lower panels we assumed that the continuum spectrum is a
black-body at the effective temperature of the central shell (1500 K - panel b)
and of the continuum photosphere (\ie\ of the region where the radial
outward optical depth is $\sim 1$). Such temperature remains
roughly constant at $\simeq 2300-2400$ K from the birth of the continuum
photosphere to the end of our calculations.

In appendix B we give a full listing of the luminosities of all the
calculated lines, plus the integrated continuum luminosity at 4 stages
(the 3 depicted in fig. \ref{spectra} plus the stage when $T_{\rm
c}=1000\;{\rm K}$). We note that the durations of the various stages are
very different: line luminosities resemble those listed in the $L_{\rm
650}$ column (corresponding to fig. \ref{spectra}.a) for $\sim10^4\;{\rm
yr}$, while they resemble those given in columns $L_{1000}$, $L_{1500}$
and $L_{\tau_{\rm c}=1}$ only for $\sim 300\;{\rm yr}$, $\sim 40\;{\rm
yr}$ and $\sim 2\;{\rm yr}$, respectively\footnote{The last duration
($\sim 2\;{\rm yr}$) is actually only a lower limit, since line emission
is still at high levels when our calculations stop; however, the \HH
envelope where line emission originates is not likely to last for $\gsim
100\;{\rm yr}$ after core formation, since it could be accreted onto
the core, or photodissociated because of the outgoing radiation.}.

We note that the spectra are produced only in the central collapsed
region: line emission comes essentially from the molecular envelope,
while continuum emission is produced close to the surface of the
hydrostatic core. In practice, the whole protostellar luminosity comes
from the central $\sim 3\msun$. The structure of this central region is
quite similar in all the models, since it is largely determined by the
self-similar collapse, rather than by initial conditions. For this
reason, the spectra emitted by different clouds during these phases
should be similar. 

\subsubsection{Radiation pressure effects}

We have found that the role of radiative force is negligible during all
the evolution we followed, \eg\ at least until the first phases of
accretion (our stage 9).
In a test run  we have 
removed  the radiative term in the momentum equation
and re-calculated evolution. We then compared the results with
those of the ``full'' code, and found that the difference is completely
negligible.

Returning to the early phase of accretion, despite the very
high luminosity ($\sim10^{36}\;\ergs$ in the continuum, plus other
$\sim10^{36}\;\ergs$ in \HH lines), which is comparable to the
Eddington luminosity $L_{\rm Edd}\simeq
1.25\times10^{36}[M/(0.01\msun)]\;\ergs$, the opacity of the infalling
gas is much smaller than the electron scattering opacity ($\kappa_{\rm
es}=\sigma_{\rm T}/m_{\rm H}\simeq 0.4\;{\rm cm^2\;g^{-1}}$), since the
infalling material is essentially molecular, with a very low ionized
fraction, and other opacity sources (Rayleigh scattering, bound-free and
free-free transitions, collision-induced absorption of \HH) give small
contributions. We find that the opacity is everywhere less than
$0.05\kappa_{\rm es}$ at all times before the formation of the
hydrostatic core; typical values are actually several order of magnitude
less than this upper limits. The same is true also after the formation
of the hydrostatic core, if we consider only the accreting layers.

\subsubsection{Bridging collapse models with accretion models}

SPS86 and recently Omukai \& Palla (2001) studied the
subsequent phase of accretion by modeling the hydrostatic core as a star
of increasing mass, surrounded by an infalling envelope whose structure
was determined by an assumed steady mass accretion rate
$\dot{M_*}=4.4\times10^{-3}\;\msun\,{\rm yr^{-1}}$.
They found that the shock surface is inside
an opticaly thick ionized region, whose outer edge is roughly coincident
with the photosphere. They also found that between the shock front and the
photosphere the radiative force term is important, since
$\kappa\simeq\kappa_{\rm es}$ and the luminosity is not far from the the
Eddington value.
Despite this fact, both works find that the radiative force is not able
to stop the accretion before the core mass grows to very high values
($>100\;\msun$).

We find that also in our models the shock front is inside the optically
thick region, but the gas between the shock and the photosphere is not
ionized, mainly because the density in the region is relatively high
($\sim 10^{-6}\;{\rm g\;cm^{-3}}$), about $1000$ times larger than the
value that can be estimated from Fig. 12 of SPS86; instead, we find that
the temperature is only slightly lower than in the SPS86 model. The
ionization fraction inside such region is low and remains such
($\lsim0.05$) even if we make the hypotesis that radiation coming from
the shock heats the gas inside the photospheric shells, so that gas and
radiation temperature are the same ($\simeq 12000\;{\rm K}$, the
post-shock temperature found both by ours and SPS86 calculations).  So,
this gas still has a low opacity, and the radiative force is small.

Although numerical effects could play some role in this discrepancy, the
most likely explanation is that the models of SPS86 are admittedly
indicative for core masses $\lsim0.1\;\msun$, \ie\ during the ``decay of
transients'' phase, when the arbitrarily chosen initial conditions are
still important; instead in our models the core mass is always in this
range (here we do not consider the simplified calculations aimed to the
study the growth of the core mass). It is quite possible that both
results are correct and that a transition occurs after the last stage of
our calculations. For instance, we note that the difference in the
density of the photospheric layers (which keeps the ionization fraction
at low levels) is mainly due to the large difference (a factor $\sim
100$) between the accretion rate we find and that assumed by SPS86;
since in our models we also find that the mass accretion rate is
decreasing, such difference will reduce.



\subsection{Comparison with metal-rich protostars}

In this section we discuss the similarities and differences between the
PopIII (high redshift, zero metal) star formation process we addressed
and the PopI-PopII (low redshift, solar metallicity) star formation.

\subsubsection{Hydrostatic core}

In our work (and also in that of ON98) the birth of the
quasi-hydrostatic core is quite different to that described by L69 (and
more recently by Masunaga, Miyama \& Inutsuka, 1998 and by Masunaga \&
Inutsuka, 2000) for a present-day protostar because of the very
different conditions.  In fact, those authors found that a molecular
core forms first, and only later on, \HH dissociation leads to the
formation of a second more internal core, which is now ionized and at
much higher temperature and density, as can be seen in table
\ref{core_comp}. The two cores coexist for a short time, while the
second core ``absorbs'' the first one.

\begin{table}
\caption{Comparison of quasi-hydrostatic cores initial properties}
\label{core_comp}
\begin{tabular}{lccc}
 &P100 & L69(I)& L69(II)\\
\hline
M/($10^{30}$ g)       & $6$          & $10$  & $3$\\
R/($10^{11}$ cm)    & $1$          & $600$ & $9$\\
T$_{\rm c}$/K            & $2\times10^4$ & $200$ & $2\times10^4$\\
$\rho_{\rm c}$/(g cm$^{-3}$)    & $1\times10^{-2}$ & $2\times10^{-10}$ &
$2\times10^{-2}$\\
f$_{\rm H_2}$ & $\lsim0.03$                    & $\simeq 1$           &
$\simeq 0$\\
\hline
\end{tabular}

\medskip
P100: results of our run P100; L69(I): first core of Larson, 1969;
L69(II): second core of Larson, 1969.
\end{table}

Instead in our models we find a single core, completely different from
the first (molecular) core of metal-rich models, but closely resembling
the second one.

The reason of both the differences and the similarities between the two
cases is essentially explained by an argument due to Omukai (2000).
He found that the locus of points in the $n-T$ plane where $\tau\sim1$,
$M\sim M_{\rm Jeans}(T,n)$ and $\Lambda_{\rm rad}\sim\Gamma_{\rm ad}$
(cfr. Silk, 1977) is approximately described by a line (L) with
$T\propto n^{2/5}$, which is shown in Fig. \ref{t_rho}; on the right of
L, a core with mass $\geq M_{\rm Jeans}(T,n)$ is optically thick.

Our metal free models become optically thick at point A
(remarkably close to L), where $T_{\rm c}$ is high and \HH
dissociation is immediately effective, preventing the formation of a
hydrostatic core.
Instead, metallic protostars ($Z\gsim10^{-6}Z_\odot$)
reach L at temperatures too low for \HH dissociation,
and the first (molecular) core forms. Such core evolves
adiabatically, \ie along a $T\propto n^{\gamma_{\rm ad}-1}$ line, which
remains close to L, because $\gamma_{\rm ad}\sim\gamma_{\rm
ad,H_2}\simeq7/5$. In this way cores of all metallicities converge
towards the vicinity of point A.
After point A, the evolution of the central regions is essentially
determined by the balance between compressional heating and \HH
dissociation cooling and no more depends on metal content; thus, metal
rich and metal free models form a core when \HH is exhausted, close to
point B.

\begin{figure}
\psfig{figure=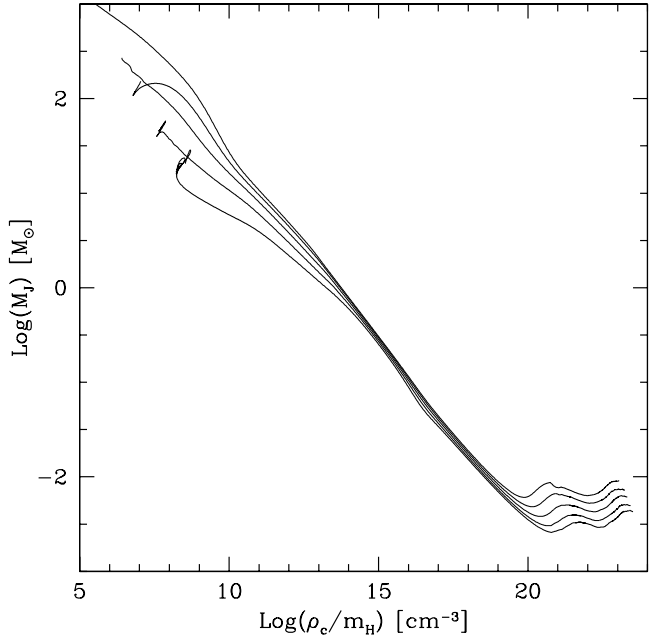,height=9.cm}
\caption{Evolution of the central Jeans mass ({\it i.e.} the Jeans mass
calculated assuming $\rho=\rho_{\rm c}$, $T=T_{\rm c}$) for several
computed models (from top to bottom: P1000, I1000, P100, I100 and
P10). Models P100/2 and N1500 are not shown but they are almost
indistinguishable from model P100.}
\label{jeansmass}
\end{figure}

The initial core mass $\sim 3\times10^{-3}\;\msun$ (only slightly dependent
upon initial conditions) agrees with $M_{\rm Jeans}(T_{\rm B},n_{\rm
B})$ (where subscript $B$ denotes the values at point B). 
Such mass is also close to the minimum value
reached by $M_{\rm
Jeans}(T_{\rm c},n_{\rm c})$ during the cloud evolution (see
Fig. \ref{jeansmass}). This minimum Jeans mass is
generally interpreted as the minimum fragment mass (\ie the mass below
which further fragmentation is impossible) and we find it depends very
weakly upon initial conditions and on metallicity. An important
corollary is that the minimum fragment mass can not be simply
found imposing  $\tau\gsim1$, since we have seen that such
condition, although necessary, is not sufficient for the formation of a
stable hydrostatic core.

It is also interesting to note that in our models there is no sign
of instability inside the hydrostatic core due, for example, to H
ionization, even if the core is mostly atomic at stage 7 and
mostlyly ionized at stage 9.

\subsubsection{Accretion rate}
Even if important core properties, such as the initial mass and the
evolutionary path of central regions in the $n-T$ plane are almost
independent of metal content, the mass accretion rates 
are completely different, with metal-free objects accreting at
a rate $\gsim10^3$ times larger than their metal enriched counterparts.
For example, in our P100 model core mass reaches $\sim 0.3 \msun$ in
$\sim 6$ years (see fig. \ref{core_mass}) while in the model of Masunaga \&
Inutsuka (2000) about $20000$ years are required for the accretion of the same
mass (see their Fig. 7).

As already discussed by several authors (\eg Stahler \etal 1980;
SPS86), the reason for this difference is
found in the initial conditions, in particular in the higher
temperature, since for an object with $M\sim M_{\rm Jeans}$ we have

\begin{equation}
\label{sst}
\dot{M}\sim (k/m_H)^{3/2}T^{3/2}/G \sim
1.8\times10^{-7} T^{3/2} \msun\,{\rm yr}^{-1}.
\end{equation}
If we take a typical initial temperature of our models $\sim300$ K we
obtain $\dot{M}\sim10^{-3}\msun\,{\rm yr}^{-1}$, but if we take $T\sim5$
K, as appropriate for present-day protostellar clouds, we get a value
$\sim500$ times smaller for $\dot{M}$.

We note that (\ref{sst}) gives the \emph{mean} accretion rate
for the \emph{complete} accretion of the protostellar cloud onto the
core; should we consider  accretion of only  a fraction of the cloud,
equation (\ref{sst}) would give higher values of $\dot{M}$ because the
appropriate reference temperature would be higher.

\subsubsection{Radiation feedback}
In the previous sections we showed that radiation pressure is not
effective in stopping the accretion flow upon a zero-metallicity
protostellar core.
The case of metal rich protostars is different: luminosities are smaller
because of smaller accretion rates, but the presence of dust brings the
opacity at much higher levels. For example, Stahler, Shu \& Taam (1981)
estimate that at the start of accretion radiative deceleration retards
the inflow of 42\% from the free fall value, even if it can not
completely stop accretion (see also the models of Masunaga \etal 1998
and Masunaga \& Inutsuka 2000, which include radiation
pressure effects in their momentum equation). In a different context,
radiative forces upon dust were also found to act decisively in halting
the accretion inflow upon massive stars (Wolfire \& Cassinelli, 1987;
Larson \& Starrfield 1971).
 
For sufficiently high density and temperature, ignition of nuclear
reactions occurs in the core interior.  To check if the effect can, in
principle, stop the accretion phase, we compare the accretion time-scale
with the diffusion time-scale of the energy produced.

As can be seen in Fig. \ref{tau_m}, the interior of the core has optical
depth $\tau_*\gsim 10^{12}$. If energy is transported radiatively, the
time-scale for its diffusion through the core is
$t_{\rm diff}\simeq{{\tau_*R_*}/{c}}\simeq 10^5\;{\rm yr}
[{{\tau_*}/{10^{12}}}] [{{R_*}/{(10^{11}{\rm cm})}}]$, where the core
radius $R_*$ has a value of the order of $5\times10^{11}$ cm. We note
that, as $T_{\rm c}$ rises, the opacity $\kappa$ inside the core
decreases; anyway, $\kappa\gsim\kappa_{\rm es}$, which leads to a strong
lower limit $\tau_*\gsim 10^{10}$.

An identical calculation for present-day protostellar cores leads to a
time-scale of the same order of magnitude.  Instead, the accretion times
differ enormously.  For example, if we extrapolate equation
(\ref{mcore_diffsol}), we find that our core would take only
$t_{1\odot}\sim 40$ years after formation to reach a mass of
$M_* \simeq 1\msun$ (but we note that $t_{1\odot}\sim250\;{\rm yr}$ also
in the models of SPS86), while in the model by Masunaga \& Inutsuka
(2000) more than $t_{1\odot}\sim1.4\times10^5$ years are required for
the core to reach the same mass.

So, $t_{1\odot}/t_{\rm diff}\sim1$ in metal-rich cores, but
$t_{1\odot}/t_{\rm diff}\ll 1$ in the metal-free case.  Accordingly, in
the metal-free case the energy generated in the core centre (for instance,
because of ignition of nuclear reactions) cannot affect the accretion,
while in the metal-rich case such energy could be able to reach the
surface and change the conditions of the infalling envelope.


The arguments discussed in the present section are subject to important
uncertainties. For example, we did not investigate any energy transport
mechanisms, such as convection, different from radiative diffusion,
which could be much more efficient (\eg, see the ``luminosity wave''
described in SPS86). 


\section{Line flux and detectability}

The work by KS01 has addresses the important problem of detectability of
the emission from collapsing primordial protostar. They employed
approximate analytical calculations (partially derived from the ON98
work) in order to estimate order-of-magnitude luminosities of collapsing
objects, assuming a central temperature $T\sim1000\;{\rm K}$.

In table \ref{spectra_comp} we list the luminosities of 5 important
lines, at all the four stages we tabulated in table \ref{line_list},
plus the total line luminosity, the continuum luminosity, and their sum.
They are compared to the values predicted by KS01.

It can be seen that KS01 predict a total luminosity comparable to our
results for the initial phases of collapse; instead, in the last phases
our total luminosities are $\gsim 10$ times larger than the KS01
value. This is quite reasonable, since the approximations they used are
much better at the beginning of the collapse rather than at the end.  If
we consider the luminosities of single lines, a more important
discrepancy emerges. KS01 correctly predict that the rotational
transition at $\lambda=9.69\mu{\rm m}$ is the most important one at the
beginning, but they estimate a very high line luminosity, that no lines
actually achieve, even in advanced stages. The reason is that they
attribute too large a fraction of the total luminosity to single
lines. Instead, the emission is actually split into a growing number of
lines: in the early phases, the two rotational lines listed in table
\ref{spectra_comp} accounts for $\sim 50\%$ of total emission (first
tabulated stage), but the luminosities of single lines ``saturate'' when
their central optical depth becomes $\gsim 1$. As the total line
emission rises, we have mainly an increase in the number of important lines,
rather than a growth of the luminosity of single lines: at the last stage no
line accounts for more than $\sim5\%$ of total line emission.

In table \ref{spectra_comp} we also give the expected specific fluxes
coming from a single protostellar object at a redshift $z=19$ in an SCDM
cosmological model at the first and the, for the first and the last
listed stage.  It can be seen that the most promising lines are those
produced by purely rotational transitions, such as those at
$\lambda=9.69\;\mu{\rm m}$ and $\lambda=8.06\;\mu{\rm m}$ (which are
redshifted to $\lambda_{\rm obs}\simeq190$ and $\lambda_{\rm
obs}\simeq160\;\mu{\rm m}$, respectively), whose expected specific flux
is $\gsim 10\;{\rm nJy}$ for at least $\sim
10^4\;{\rm yr}$); vibrational transitions, such as those at
$\lambda=2.41\;\mu{\rm m}$ and $\lambda=2.81\;\mu{\rm m}$ (redshifted to
$\lambda_{\rm obs}\simeq48\;\mu{\rm m}$ and $\lambda_{\rm
obs}\simeq56\;\mu{\rm m}$) reach specific fluxes of the same order only
in the last phases and their emission at such levels should last only
for $\lsim10^3\;{\rm yr}$. Finally, we predict that the continuum
specific flux should be of the order of $0.002\;{\rm nJy}$ at an
observed wavelength of $\lambda_{obs}\sim 40\mu{\rm m}$ (from a
rest-frame wavelength of $\sim 2\mu{\rm m}$); we cannot predict the
duration of continuum emission, since it is deeply affected by the
evolution of the accretion and by the evolution of the hydrostatic core
at the centre.

Such fluxes are largely below the detection threshold of any
observational facility that will be available in the near future.
NGST sensitivity for continuum radiation should be of the
order of $\sim1\;\mu{\rm Jy}$ at $\lambda=20\;\mu{\rm m}$, while ALMA
should reach a sensitivity of $\sim 10\;{\rm mJy}$ for a line $25\;{\rm
km\,s^{-1}}$ wide at $\lambda\simeq350\;\mu{\rm m}$. However, the
possibility that primordial protostars form in large groups must not to
be excluded, since a galactic-scale ($10^{11}\;\msun$) fluctuation could
host a large number of $10^6\;\msun$ halos, and each of them could host
several protostellar objects. For instance KS01 roughly estimate that a
galactic-scale fluctuation could have a luminosity $\sim 10^4$ times
larger than a single object, and that such luminosity could last for
$\sim 10^7\;{\rm yr}$.  If this is correct, specific fluxes from
galactic-scale objects should be of the order of $\sim 0.01\;\mu{\rm
Jy}$ and $\sim 0.1\;{\rm mJy}$ for continuum and line emission,
respectively.





\begin{table}
\caption{Comparison of line and continuum luminosities; luminosities (L)
are in units of $10^{34}\;{\rm erg\,s^{-1}}$, specific fluxes at earth
(l, for an object at $z=19$ in a SCDM cosmology with $H_0=75\;{\rm
km\,s^{-1}\,Mpc^{-1}}$) in nJy; the continuum specific flux is
calculated at the peak of the black-body spectrum.
($\lambda\simeq2\mu{\rm m}$). All data come from run P100/2}
\label{spectra_comp}
\begin{tabular}{cccccccc}
 $\lambda[\mu{\rm m}]$ & $L_{650}$ & $L_{1000}$ & $L_{1500}$ &
$L_{\tau_{\rm c}=1}$ & $L(KS01)$ & $l_{650}$  &$l_{\tau_{\rm c}=1}$\\
\hline
2.41      & 0.20 & 2.21 & 3.90 & 4.26 & 0.4  & 0.6 & 13\\
2.81      & 0.18 & 2.49 & 4.22 & 4.55 & -    & 0.7 & 16\\
8.06      & 0.99 & 3.48 & 3.72 & 3.74 & -    & 10  & 37\\
9.69      & 1.12 & 2.78 & 2.90 & 2.90 & 5.5  & 14  & 35\\
28.3      & 0.01 & 0.03 & 0.03 & 0.03 & 0.8  & 0.3 & 1.0\\
Line total& 4.4  & 43   & 89   & 106  & 6.7  & -   & -   \\
Continuum & -    & 0.00 & 0.13 & 75   & -    & -   & 0.002  \\
Total     & 4.4  & 43.2 & 89.3 & 181  & 6.7  & -   & -   \\
\hline
\end{tabular}
\end{table}
\medskip

\section{Conclusions}                 
In this work we have investigated the evolution of the luminosity 
and spectra of the radiation emitted by primordial collapsing
protostars. We find that initially radiation comes mostly from \HH
rotational lines. Later, the contributions from \HH
vibrational lines and also from continuum radiation (due to the accretion
upon a centrally hydrostatic core) become important. At early
stages, only two rotational lines account for about 50\%\ of the total
emission, but the subsequent increase in total line
luminosity results in an increase of the number of prominent lines,
rather than a large increase in the luminosity of the most important
ones. As a result, the expected line specific fluxes are always $\lsim
50\;{\rm nJy}$ (with line widths $\lsim 30\;{\rm km\,s^{-1}}$), even for
the most important lines, which should fall at observed wavelengths in
the range $150-200\mu{\rm m}$. Continuum radiation should reach a
specific flux of the order of $0.002\;{\rm nJ}$, peaking at an observed
wavelength between $20$ and $40\;\mu\rm m$. These fluxes should be
quite universal, provided that a molecular central region of $\sim
3\msun$ develops at the centre of each protostellar cloud. 
The predicted fluxes, both for line and continuum radiation, are
largely below the detection threshold of existing or planned instruments
(such as NGST and ALMA), but the possibility that primordial
protostars form large ``clusters'' remains to be investigated, even if
rough estimates do not look very promising (KS01).

In our work we also found that, different from the present-day case,
radiation forces play virtually no role in the initial evolution of a
protostellar object (\ie, during the self-similar collapse, the
formation of the core and the onset of accretion). Although we were not
able to completely bridge the gap with accretion models such as those of
SPS86 and Omukai \& Palla 2001, we have shown that their initial
assumptions are quite reasonable. Even if there are differences in the
ionization structure, there is indication (in particular, the decrease
in the mass accretion rate) that these will vanish as the hydrostatic
core mass grows.  Thus, undisturbed spherical collapse should proceed
until the initial protostellar cloud has been largely accreted, or until
radiation forces halt the infall at $M_*\sim 300\;\msun$ (Omukai \&
Palla 2001).

What can stop mass growth is the presence of HD molecules (even in a
tiny fraction) that may result from shocks (Flower \& Pineau des
For\^ets 2001, Uehara \& Inutsuka 2000); in this case the fragmentation
of the PopIII object might end up in clumps of roughly solar rather than
thousand solar masses, thus drastically changing the final mass of the
star.

If this process is not important and angular momentum and magnetic
fields do not play a significant role either, the final mass of a
primordial protostar is primarily set by the amount of gas available in
the parent cloud. Then, it is conceivable that the primordial IMF is
skewed towards high masses, and that very massive stars (VMS),
with masses in excess of $M\sim 100\;\msun$, can form. Such
stars may end-up as a population of intermediate mass black holes
(Portinari \etal 1998, Heger \& Langer 2000) that represent the end
product of such an episode of pregalactic star formation. Then, these
seeds could be incorporated into the larger units as seen in the bulges
of present-day galaxies (Madau \& Rees 2001).

\medskip

We thank an anonymous referee for a critical reading and of the
manuscript and for stimulating remarks. We also thank B. Ciardi for
discussion and for making her chemistry routines available to us, and
K. Omukai, R. Nishi and F. Palla for discussion and clarifying remarks
about their works.


\vskip 1truecm

\appendix
\section{The code}

Hereafter we describe the main features of the code we developed.


\subsection{Generalities}

\subsubsection{Gas hydrodynamics}

Despite the cosmological context of our calculations, the collapsing
baryon clouds we are studying are assumed to be self-gravitating, and we
can neglect the gravitational effects of host dark matter halo.

The code integrates the following 1-D Lagrangian hydrodynamical
equations (conservation of mass, momentum and energy) using an explicit
finite-difference scheme:

\begin{equation}
\label{continuity}
{{dM_r}\over{dr}} = 4 \pi r^2 \rho
\end{equation}
\begin{equation}
\label{momentum}
{{dv}\over{dt}} = -4 \pi r^2 {{dp}\over{dM_r}} - {{GM_r}\over{r^2}} +
f_{\rm rad}
\end{equation}
\begin{equation}
\label{energy}
{{du}\over{dt}} = - p {{d}\over{dt}} \left({1\over\rho}\right) +
{{\Gamma-\Lambda}\over\rho},
\end{equation}
where the Lagrangian coordinate $M_r$ is the mass inside a radius $r$.
$\rho$, $v$, $p$, $f_{\rm rad}$, $u$, represent the density, velocity,
pressure, radiative force per unit mass and internal energy per unit
mass, respectively. $\Gamma$ and $\Lambda$ are the heating and cooling
rate per unit volume, respectively.

Hydrodynamical effects are included by adding to the pressure an
artificial viscosity term of the form
\begin{equation}
\label{art_visc}
q\propto \rho\, {\Delta v}^2,
\end{equation}
in regions where contiguous gas shells would compenetrate (\eg Bowers \&
Wilson 1991; Thoul \& Weinberg 1995; hereafter, TW95).


\subsubsection{Chemistry}
Because of the extremely low metallicity of the primordial material as
predicted by standard big-bang nucleosynthesis ($Z\sim10^{-12}$; Kolb \&
Turner 1990), we can safely limit our treatment of chemistry to the
various ionic species of atomic hydrogen (H$^0$, H$^+$, H$^-$),
molecular hydrogen (\HH, \HHP), and helium (He$^0$, He$^+$, He$^{++}$),
plus free electrons (e$^-$).

The He number abundance is taken to be 1/12 of the total H number
abundance; this translates into mass fractions of X=0.75, Y=0.25 and
Z=0.

Some authors (Flower 2000; Flower \& Pineau de For\^ets 2001; and also,
in a different context, Uehara \& Inutsuka 2000) suggested that
deuterium is an important coolant in a metal-free gas, because of the
permanent dipole moment of HD molecules and their low excitation
temperature ($\sim 160$ K for HD instead of $\sim510$ K for
\HH). However HD cooling is likely to be important only at temperatures
below some hundred degrees, \ie during the formation of protostellar
clouds, rather than during their collapse. For this reason, we neglect
the effects of deuterium.  Note that BCL01 included HD cooling in their
investigation of primordial cloud fragmentation, and found that (for a
standard Deuterium abundance) the thermal evolution of the gas would not
change significatively.


Initial conditions were chosen with chemical abundances far from
equilibrium, and we follow their evolution by integrating a network of
reactions (see Table 1). In particular, we include the 3-body reactions
described by Palla, Salpeter \& Stahler (1983).

\begin{table}
\caption{Chemical reactions and reaction rates}
\label{chem_reac}
\begin{tabular}{rlc}
\multicolumn{2}{c}{Reaction}& Rate\\
\hline
H$^+$ + e$^-$ $\to$ & H + $\gamma$          &
AANZ98, $k_2$\\
H + e$^-$     $\to$ & H$^-$ + $\gamma$      &
AANZ98, $k_7$\\
H + H$^-$     $\to$ & \HH + e$^-$         &
AANZ98, $k_8$\\
H + H + H     $\to$ & \HH + H             &
PSS83, $k_4$\\
\HH + H     $\to$ & H + H + H             &
PSS83, $k_5$\\
H + H + \HH $\to$ & \HH + \HH         &
PSS83, $k_6$\\
\HH + \HH $\to$ & H + H + \HH         &
PSS83, $k_7$\\
H + e$^-$     $\to$ & H$^+$ + e$^-$ + e$^-$ &
AANZ98, $k_1$\\
H + H         $\to$ & H$^+$ + H + e$^-$     &
PSS83, $k_9$\\
\hline
\end{tabular}

\medskip
AANZ98 stands for Abel, Anninos, Norman \& Zhang (1998) and PSS83 stands
for Palla, Salpeter \& Stahler (1983). Note that the H recombination
rate (first reaction in the list) is appropriate for case A (case B
could be more appropriate, but results are not affected by the
difference, as can be seen from the comparison to ON98).
\end{table}

As the gas gets denser ($\rho\gsim \rho_{\rm eq}\equiv 5 \times
10^{-11}$ \gcc), the abundances of the relevant species (\HH, H$^0$ and
He$^0$, which then represent more than 99.9\% of the total number
abundance) are close to their equilibrium values, as reactions. We then
switch to the equilibrium chemistry.

When the gas is almost ideal, \ie\ when $p\leq10^4\;{\rm dyne\,cm^{-2}}$
($p$ is the gas pressure) we find the equilibrium abundances by solving
the Saha equations for the relevant species. However these equations
become inappropriate at high temperatures and densities, so when
$p\geq10^6\;{\rm dyne\,cm^{-2}}$ we interpolate the abundance values
from the tables given by SCVH95. In the intermediate regime ($10^4\;{\rm
dyne\,cm^{-3}}\leq p \leq10^6\;{\rm dyne\,cm^{-2}}$), we evaluate the
chemical composition both with Saha equations and using the SCVH95
tables, and then we appropriately interpolate the two results; this is
done in order to ensure a smooth transition between the two regimes.


\subsubsection{Thermodynamical properties}
For the solution of the hydrodynamical equations, we need to specify the
equation of state and to calculate the internal energy as a function of
temperature, density and chemical composition.

At low densities and temperatures (in practice, we use $p\leq10^4\;{\rm
dyne\,cm^{-2}}$ as a criterion), the gas is ideal to a very good
level of approximation and we can use the usual equation
\begin{equation}
\label{ideal_eos}
p=n k T
\end{equation}
where $n$ is the number density of the gas.

In this ideal regime we can also write the total internal energy per
unit mass as
\begin{equation}
\label{ideal_u}
u=u_{\rm th}+u_{\rm chem}
\end{equation}
with
\begin{equation}
\label{ideal_uth}
u_{\rm th}={1\over{\gamma_{\rm ad}-1}}N_{\rm tot}kT,\qquad
\end{equation}
\begin{equation}
\label{ideal_uchem}
u_{\rm chem}=-N_{\rm tot}\sum_{i=1}^{9}{y_i\chi_i}
\end{equation}
where $u_{\rm th}$ and $u_{\rm chem}$ are the thermal and the chemical
energy per unit mass, while $N_{\rm tot}$, $y_i$ and $\chi_i$ are the
total number of particles per unit mass, the number
fraction of particles of species $i$ ($y_i=N_i/N_{\rm tot}$), and the
binding energy of a particle of species $i$, respectively. Last,
$\gamma_{\rm ad}$ is the mean adiabatic index of the gas, obtained by a
weighted average of the adiabatic indexes of the various chemical
species: for the seven included mono-atomic species we have $\gamma_{\rm
ad,X}=5/3$, while for the two diatomic species we use $\gamma_{\rm
ad,H_2^+}=7/5$ and the expression (from ON98) is
\begin{equation}
\label{gammah2}
{{1}\over{\gamma_{ad,H_2}-1}}={5\over2}+x^2{{e^x}\over{(e^x-1)^2}},\qquad
x={{6100 K}\over{T}}.
\end{equation}
The value of $\gamma$ for H$_2^+$ neglects the vibrational degrees of freedom, but the
abundance of this ion is always very small and this does not
introduce any significant error.



When the gas is no more ideal ($p\geq10^6\;{\rm dyne\,cm^{-2}}$), we find
$p(\rho,T)$ and $u(\rho,T)$ by interpolating the values tabulated by
SCVH95, which have been calculated jeeping non-ideal effects into
account. In the transition between the two regimes, we interpolate the
results of the ideal and of the SCVH95 values in order to obtain a smooth
transition.

\subsubsection{Radiative transfer and cooling}
In the initial stages of collapse, gas cooling is dominated by the \HH
roto-vibrational transitions, which soon become optically thick. A
careful treatment of radiative transfer effects is then crucial in the
study of the collapse.

\HH line cooling is important only when $T\lsim 2500$ K, as at higher
temperatures \HH molecules slowly dissociate, and continuum emission
starts dominating the local cooling.  The energy gaps between \HH rotational
levels typically correspond to excitation temperatures of
few$\times100$ K - $10^3$ K, while the gap between vibrational levels
corresponds to $\sim 6000$ K. Under such conditions, it is adequate to
consider only \HH roto-vibrational levels with rotational quantum number
$0\leq J\leq20$, and vibrational quantum number $0\leq v\leq2$. In this
way, all the states whose excitation temperatures (relative to the
ground state) is $\lsim 20000$ K are considered. We treat the radiative
transfer of all the lines (between the considered levels) permitted by
the quadrupole selection rules.

Line opacities are computed using the Einstein A-coefficients from
Turner, Kirby-Docken \& Dalgarno (1977), and assuming local
thermodynamic equilibrium (LTE) for estimating the level
populations. This is correct since the density at which a line gives a
relevant contribution to the total cooling is always greater than the
critical density at which collisional de-excitations dominate upon
radiative transitions.

Finally, the effects of thermal broadening and Doppler shifts due to
bulk motions arising in the collapse phase have been included in the
line transfer calculations.

When the temperature rises to a sufficiently high value ($T>1000$ K), we
include the cooling by continuum radiation; continuum radiative transfer
is treated as a gray problem, using the mean Planck opacity for $Z=0$
gas from Lenzuni, Chernoff \& Salpeter (1991) when $1000\;K<T<7000$
K, and the Rosseland mean opacity for metal-free gas from Rogers \&
Iglesias (1992) when $T>7000$ K.

\subsection{Numerical details}

\subsubsection{Implementation of the basic equations}

The numerical treatment of the dynamical evolution is made along the
lines described by Bowers \& Wilson (1991), and TW95.

With the exception of the energy equation, the dynamical equations
listed in \S 2.1 were discretized as in TW95
(their equations 9, 10, 11, 12, 13 and 15), assuming a value of 4 for the
artificial viscosity parameter $c_q$.

Such Lagrangian finite difference scheme is second order accurate in
space and time.

For the energy equation (\ref{energy}), our treatment starts from
equation (14) of TW95. 
The chemical term in the internal energy (see \S 2.3) introduces a complex
dependence upon the final temperature. For
this reason the energy equation is solved iteratively, using the bisection
method. In this case the result is only first order accurate in time.

\subsubsection{Time-steps}
The integration time-step is chosen evaluating the minimum
among:
\begin{itemize}
\item[-]{twice the previous integration time-step;}
\item[-]{the minimum dynamical time-scale over all shells $i$
($\min_i\{C_{\rm dyn}({Gr_i^3/M_{r,i}})^{1/2}\}$);}
\item[-]{the minimum Courant time-scale over all shells $i$
($\min_i\{C_{\rm Cour}\Delta r_i/c_{{\rm s},i}\}$);}
\item[-]{the minimum cooling time-scale over all shells $i$
($\min_i\{C_{\rm cool}u_i\rho/(\Gamma_i-\Lambda_i)\}$);}
\item[-]{the minimum crossing time-scale over all shells $i$
($\min_i\{C_{\rm cross}\Delta r_i/\Delta v_i\}$);}
\end{itemize}
In addition, we check \emph{a posteriori} that the temperature 
does not change by more than 0.5\%\ in a single time-step.
The factors $C_{\rm dyn}=1$, $C_{\rm Cour}=0.2$, $C_{\rm cool}=0.1$
and $C_{\rm cross}=0.05$ are safety constants.

\subsubsection{Boundary conditions}

As internal boundary condition we simply require that the
centre of our object is a geometrical point ($r=0$, $M_r=M_0=0$), with 
zero velocity ($v_{\rm in}=0$).

The external boundary condition is set by taking a constant pressure outside
the most external shell.

\subsubsection{Implementation of radiative transfer}

{\bf General scheme}

The solution of the radiative transfer equation in spherical symmetry
requires to find the specific intensity
$I_\nu(r,\mu)$, where $\mu$ is the cosine of the angle $\theta$ between the
radial outward direction and the direction of the considered ray.

This is done by means
of the standard tangent ray method (Hummer \& Rybicki 1971; Bowers \&
Wilson 1991; ON98).  In this scheme, the specific intensity
$I_\nu(r,\mu)$ is evaluated at the external radius $r_i$ of each shell
for $1+2i$ values of $\mu$.

Once $I(r_i,\mu_j)$ is known, we compute the monochromatic flux 
and the net outward bolometric luminosity $L(r_i)$ at each shell radius. 

The cooling rate (per unit time per unit volume) of the shell $i$ is then
given by
\begin{equation}
\label{radiative_cooling}
\Lambda_{rad,i}={{L(r_i)-L(r_{i-1})}\over{(4\pi/3)(r_i^3-r_{i-1}^3)}}
\end{equation}
\smallskip
{\bf \HH lines}

To calculate the source function and the absorption coefficient,
we assume LTE. This is a fair assumption, as the  
density is always much greater than the relevant critical values.

The absorption coefficient is given by 
\begin{equation}
\label{absorpt_general}
\alpha_\nu={{h\nu}\over{4\pi}}\phi(\nu)
\left({{{2h\nu^3}\over{c^2}}A_{21}}\right)
\left({n_1{{g_2}\over{g_1}}-n_2}\right)
\end{equation}
where $\phi(\nu)$ is the line profile, $A_{21}$ is the
Einstein coefficient for the relevant transition (taken from Turner
\etal 1977), $g_2$ and $g_1$ are the statistical weights of the two
levels involved in the transition, and $n_2$ and $n_1$ are number
densities of molecules in the two considered levels, computed assuming LTE.

As noted by ON98, since the temperature typical of \HH line emission is  
quite low, the line profiles are narrow, and the Doppler shift caused 
by the gas bulk infall can be important. 
We assume a Gaussian profile for each line, sampled in 30 frequency bins, and we properly 
take into account Doppler shift in the line transfer.
If the line absorption depth is $<0.1$, we use the optically thin approximation.
\newline\smallskip
{\bf Continuum}

The continuum transfer is treated as a gray problem. 
Again, we assume LTE to estimate the source function. 
The absorption coefficient is calculated from the (Planck) opacity tables 
of Lenzuni \etal 1991 when $T<7000$ K, and
from the (Rosseland) opacity tables of Rogers \& Iglesias 1992 for $T>7000$ K. 

As the continuum optical depth can reach values $\gg1$ (see fig. \ref{tau_m}),
the numerical treatment above can become inaccurate. 
For this reason, when $\tau>4$ in a single shell (see the conditions listed in
ON98), we use the diffusion approximation, \ie 
\begin{equation}
\label{flux_continuum_diff}
F=
-{{16\sigma T^3}\over{3\alpha}}{{\partial T}\over{\partial r}}\simeq
-{{16\sigma
T_i^3}\over{3\alpha_i}}{{T_{i+1}-T_{i-1}}\over{r_{i+1}-r_{i-1}}}
\end{equation}
where $\sigma$ is the Stefan-Boltzmann constant.

\section{List of emitted lines}

In the following table we list the luminosity at infinity of each \HH line we
included into our code. Lines are sorted according to their wavelength.

The columns give, respectively:
\begin{itemize}
\item[-]{the quantum levels describing the transition;}
\item[-]{the wavelength (in $\mu{\rm m}$) of the line;}
\item[-]{the radial optical depth from the centre of the simulated
object to infinity, as evaluated at line centre, and at the epoch when
the continumm optical depth is $\tau_{\rm c}=1$;}
\item[-]{the line total luminosity at infinity, evaluated at four
different epochs, \ie\ when the central temperature is 650 K ($L_{650}$),
1000 K ($L_{1000}$) and 1500 K ($L_{1500}$), and when $\tau_{\rm c}=1$
($L_{\tau_{\rm c}=1}$); such luminosities are given in ${\rm erg\;s^{-1}}$;} 
\item[-]{the specific luminosity (in ${\rm
erg\;s^{-1}\;Hz^{-1}}$) at the stage when $\tau_{\rm c}=1$,
estimated using a line width equal to the Doppler broadening $\Delta\nu
= (\nu_0/c) (2kT / \mu m_{\rm H})^{1/2}$, where $\nu_0$ is the line
central frequency, $\mu\simeq2$ is the \HH molecular weight and $T$ is the
temperature of the shell where the radial optical depth to infinity at
line centre is equal to 1 (or $T\simeq2338\;{\rm K}$ for the few lines with
$\tau<1$).}
\end{itemize}

The last lines of the table give the total line luminosities, the
continuum luminosity and the sum of the two. Continuum luminosity is not
calculated (being actually very small) when the central temperature is
$\leq 1000\;{\rm K}$; continuum specific luminosity is only a rough
estimate, since we took $\Delta\nu=\nu_{\rm peak}$, where $\nu_{\rm
peak}\simeq1.4\times10^{14}\;{\rm Hz}$ is the frequency of the peak of
blackbody emission at $T=2338\;{\rm K}$.

\begin{table*}
\caption{Line list}
\label{line_list}
\begin{tabular}{lccccccc}
\hline
$(v_{\rm i},J_{\rm i})\rightarrow(v_{\rm f},J_{\rm f})$ &
$\lambda[\mu{\rm m}]$ & $\tau$ & $L_{650}$ & $L_{1000}$ & $L_{1500}$ &
$L_{\tau_{\rm c}=1}$ & $l_{\tau_{\rm c}=1}$ \\
\hline
$(2,11)\rightarrow(0,9)$ & 1.0700 & $4.39\times10^{1}$ & $9.91\times10^{24}$ & $1.59\times10^{29}$ & $6.50\times10^{31}$ & $4.55\times10^{32}$ & $9.77\times10^{22}$ \\ 
$(2,10)\rightarrow(0,8)$ & 1.0701 & $7.65\times10^{1}$ & $7.00\times10^{25}$ & $6.32\times10^{29}$ & $1.54\times10^{32}$ & $6.50\times10^{32}$ & $1.40\times10^{23}$ \\ 
$(2,12)\rightarrow(0,10)$ & 1.0736 & $2.34\times10^{1}$ & $1.24\times10^{24}$ & $3.64\times10^{28}$ & $2.36\times10^{31}$ & $2.85\times10^{32}$ & $6.13\times10^{22}$ \\ 
$(2,9)\rightarrow(0,7)$ & 1.0737 & $1.24\times10^{2}$ & $4.32\times10^{26}$ & $2.25\times10^{30}$ & $2.97\times10^{32}$ & $9.08\times10^{32}$ & $2.26\times10^{23}$ \\ 
$(2,8)\rightarrow(0,6)$ & 1.0806 & $1.83\times10^{2}$ & $2.25\times10^{27}$ & $6.99\times10^{30}$ & $4.84\times10^{32}$ & $1.19\times10^{33}$ & $2.97\times10^{23}$ \\ 
$(2,13)\rightarrow(0,11)$ & 1.0809 & $1.18\times10^{1}$ & $1.43\times10^{23}$ & $7.73\times10^{27}$ & $8.01\times10^{30}$ & $1.76\times10^{32}$ & $3.48\times10^{22}$ \\ 
$(2,7)\rightarrow(0,5)$ & 1.0908 & $2.46\times10^{2}$ & $9.68\times10^{27}$ & $1.85\times10^{31}$ & $6.81\times10^{32}$ & $1.44\times10^{33}$ & $3.65\times10^{23}$ \\ 
$(2,14)\rightarrow(0,12)$ & 1.0925 & $5.57\times10^{0}$ & $1.55\times10^{22}$ & $1.54\times10^{27}$ & $2.58\times10^{30}$ & $1.05\times10^{32}$ & $2.09\times10^{22}$ \\ 
$(2,6)\rightarrow(0,4)$ & 1.1042 & $2.98\times10^{2}$ & $3.36\times10^{28}$ & $3.91\times10^{31}$ & $8.61\times10^{32}$ & $1.67\times10^{33}$ & $4.26\times10^{23}$ \\ 
$(2,15)\rightarrow(0,13)$ & 1.1087 & $2.51\times10^{0}$ & $1.61\times10^{21}$ & $2.97\times10^{26}$ & $7.98\times10^{29}$ & $5.79\times10^{31}$ & $1.18\times10^{22}$ \\ 
$(2,5)\rightarrow(0,3)$ & 1.1210 & $3.19\times10^{2}$ & $9.17\times10^{28}$ & $6.66\times10^{31}$ & $1.02\times10^{33}$ & $1.89\times10^{33}$ & $4.91\times10^{23}$ \\ 
$(2,16)\rightarrow(0,14)$ & 1.1300 & $1.07\times10^{0}$ & $1.65\times10^{20}$ & $5.53\times10^{25}$ & $2.38\times10^{29}$ & $2.80\times10^{31}$ & $5.79\times10^{21}$ \\ 
$(2,4)\rightarrow(0,2)$ & 1.1410 & $2.96\times10^{2}$ & $1.90\times10^{29}$ & $9.34\times10^{31}$ & $1.19\times10^{33}$ & $2.06\times10^{33}$ & $5.45\times10^{23}$ \\ 
$(2,17)\rightarrow(0,15)$ & 1.1573 & $0.43\times10^{0}$ & $1.69\times10^{19}$ & $1.00\times10^{25}$ & $6.79\times10^{28}$ & $1.19\times10^{31}$ & $2.52\times10^{21}$ \\ 
$(2,3)\rightarrow(0,1)$ & 1.1645 & $2.27\times10^{2}$ & $2.81\times10^{29}$ & $1.09\times10^{32}$ & $1.29\times10^{33}$ & $2.20\times10^{33}$ & $5.93\times10^{23}$ \\ 
$(2,2)\rightarrow(0,0)$ & 1.1914 & $1.25\times10^{2}$ & $2.59\times10^{29}$ & $9.28\times10^{31}$ & $1.24\times10^{33}$ & $2.13\times10^{33}$ & $5.88\times10^{23}$ \\ 
$(2,18)\rightarrow(0,16)$ & 1.1915 & $0.16\times10^{0}$ & $1.78\times10^{18}$ & $1.81\times10^{24}$ & $1.89\times10^{28}$ & $4.54\times10^{30}$ & $9.91\times10^{20}$ \\ 
$(2,19)\rightarrow(0,17)$ & 1.2338 & $5.57\times10^{-2}$ & $1.88\times10^{17}$ & $3.14\times10^{23}$ & $4.87\times10^{27}$ & $1.55\times10^{30}$ & $3.49\times10^{20}$ \\ 
$(2,1)\rightarrow(0,1)$ & 1.2400 & $1.20\times10^{2}$ & $3.59\times10^{29}$ & $1.17\times10^{32}$ & $1.47\times10^{33}$ & $2.44\times10^{33}$ & $7.01\times10^{23}$ \\ 
$(2,2)\rightarrow(0,2)$ & 1.2439 & $1.23\times10^{2}$ & $2.69\times10^{29}$ & $1.02\times10^{32}$ & $1.44\times10^{33}$ & $2.46\times10^{33}$ & $7.08\times10^{23}$ \\ 
$(2,3)\rightarrow(0,3)$ & 1.2498 & $1.31\times10^{2}$ & $1.78\times10^{29}$ & $8.42\times10^{31}$ & $1.37\times10^{33}$ & $2.40\times10^{33}$ & $6.94\times10^{23}$ \\ 
$(2,4)\rightarrow(0,4)$ & 1.2577 & $1.24\times10^{2}$ & $9.04\times10^{28}$ & $5.69\times10^{31}$ & $1.23\times10^{33}$ & $2.24\times10^{33}$ & $6.53\times10^{23}$ \\ 
$(2,5)\rightarrow(0,5)$ & 1.2676 & $1.06\times10^{2}$ & $3.60\times10^{28}$ & $3.13\times10^{31}$ & $1.00\times10^{33}$ & $2.01\times10^{33}$ & $5.89\times10^{23}$ \\ 
$(2,6)\rightarrow(0,6)$ & 1.2796 & $8.25\times10^{1}$ & $1.14\times10^{28}$ & $1.45\times10^{31}$ & $7.18\times10^{32}$ & $1.67\times10^{33}$ & $4.30\times10^{23}$ \\ 
$(2,20)\rightarrow(0,18)$ & 1.2860 & $1.58\times10^{-2}$ & $1.88\times10^{16}$ & $4.89\times10^{22}$ & $1.08\times10^{27}$ & $4.39\times10^{29}$ & $1.03\times10^{20}$ \\ 
$(2,7)\rightarrow(0,7)$ & 1.2938 & $6.00\times10^{1}$ & $2.99\times10^{27}$ & $5.78\times10^{30}$ & $4.51\times10^{32}$ & $1.30\times10^{33}$ & $3.37\times10^{23}$ \\ 
$(2,0)\rightarrow(0,2)$ & 1.2947 & $6.98\times10^{1}$ & $2.58\times10^{29}$ & $8.42\times10^{31}$ & $1.30\times10^{33}$ & $2.28\times10^{33}$ & $6.82\times10^{23}$ \\ 
$(2,8)\rightarrow(0,8)$ & 1.3101 & $4.04\times10^{1}$ & $6.45\times10^{26}$ & $2.01\times10^{30}$ & $2.38\times10^{32}$ & $9.45\times10^{32}$ & $2.49\times10^{23}$ \\ 
$(2,9)\rightarrow(0,9)$ & 1.3287 & $2.59\times10^{1}$ & $1.20\times10^{26}$ & $6.24\times10^{29}$ & $1.10\times10^{32}$ & $6.46\times10^{32}$ & $1.72\times10^{23}$ \\ 
$(2,1)\rightarrow(0,3)$ & 1.3372 & $8.54\times10^{1}$ & $2.76\times10^{29}$ & $9.68\times10^{31}$ & $1.52\times10^{33}$ & $2.65\times10^{33}$ & $8.19\times10^{23}$ \\ 
$(2,10)\rightarrow(0,10)$ & 1.3497 & $1.57\times10^{1}$ & $1.94\times10^{25}$ & $1.75\times10^{29}$ & $4.64\times10^{31}$ & $4.30\times10^{32}$ & $1.16\times10^{23}$ \\ 
$(2,11)\rightarrow(0,11)$ & 1.3730 & $9.12\times10^{0}$ & $2.80\times10^{24}$ & $4.50\times10^{28}$ & $1.85\times10^{31}$ & $2.68\times10^{32}$ & $6.74\times10^{22}$ \\ 
$(2,2)\rightarrow(0,4)$ & 1.3838 & $7.50\times10^{1}$ & $1.81\times10^{29}$ & $7.26\times10^{31}$ & $1.42\times10^{33}$ & $2.52\times10^{33}$ & $8.08\times10^{23}$ \\ 
$(2,12)\rightarrow(0,12)$ & 1.3988 & $5.14\times10^{0}$ & $3.72\times10^{23}$ & $1.09\times10^{28}$ & $7.05\times10^{30}$ & $1.70\times10^{32}$ & $4.35\times10^{22}$ \\ 
$(2,13)\rightarrow(0,13)$ & 1.4271 & $2.81\times10^{0}$ & $4.64\times10^{22}$ & $2.50\times10^{27}$ & $2.59\times10^{30}$ & $9.89\times10^{31}$ & $2.58\times10^{22}$ \\ 
$(2,3)\rightarrow(0,5)$ & 1.4349 & $5.51\times10^{1}$ & $8.39\times10^{28}$ & $4.11\times10^{31}$ & $1.08\times10^{33}$ & $2.13\times10^{33}$ & $6.14\times10^{23}$ \\ 
$(2,14)\rightarrow(0,14)$ & 1.4581 & $1.51\times10^{0}$ & $5.63\times10^{21}$ & $5.62\times10^{26}$ & $9.38\times10^{29}$ & $5.56\times10^{31}$ & $1.48\times10^{22}$ \\ 
$(2,4)\rightarrow(0,6)$ & 1.4903 & $3.49\times10^{1}$ & $2.88\times10^{28}$ & $1.82\times10^{31}$ & $6.56\times10^{32}$ & $1.55\times10^{33}$ & $4.65\times10^{23}$ \\ 
$(2,15)\rightarrow(0,15)$ & 1.4918 & $0.81\times10^{0}$ & $6.88\times10^{20}$ & $1.26\times10^{26}$ & $3.40\times10^{29}$ & $3.01\times10^{31}$ & $8.22\times10^{21}$ \\ 
$(2,16)\rightarrow(0,16)$ & 1.5283 & $0.44\times10^{0}$ & $8.69\times10^{19}$ & $2.90\times10^{25}$ & $1.25\times10^{29}$ & $1.59\times10^{31}$ & $4.46\times10^{21}$ \\ 
$(2,5)\rightarrow(0,7)$ & 1.5503 & $1.97\times10^{1}$ & $7.54\times10^{27}$ & $6.56\times10^{30}$ & $3.12\times10^{32}$ & $1.01\times10^{33}$ & $3.14\times10^{23}$ \\ 
$(2,17)\rightarrow(0,17)$ & 1.5676 & $0.24\times10^{0}$ & $1.17\times10^{19}$ & $6.96\times10^{24}$ & $4.72\times10^{28}$ & $8.39\times10^{30}$ & $2.41\times10^{21}$ \\ 
$(2,18)\rightarrow(0,18)$ & 1.6097 & $0.14\times10^{0}$ & $1.76\times10^{18}$ & $1.79\times10^{24}$ & $1.86\times10^{28}$ & $4.49\times10^{30}$ & $1.32\times10^{21}$ \\ 
$(2,6)\rightarrow(0,8)$ & 1.6146 & $9.92\times10^{0}$ & $1.54\times10^{27}$ & $1.94\times10^{30}$ & $1.21\times10^{32}$ & $5.81\times10^{32}$ & $1.88\times10^{23}$ \\ 
$(2,19)\rightarrow(0,19)$ & 1.6546 & $7.92\times10^{-2}$ & $3.03\times10^{17}$ & $5.08\times10^{23}$ & $7.86\times10^{27}$ & $2.50\times10^{30}$ & $7.57\times10^{20}$ \\ 
$(2,7)\rightarrow(0,9)$ & 1.6830 & $4.58\times10^{0}$ & $2.51\times10^{26}$ & $4.85\times10^{29}$ & $4.12\times10^{31}$ & $2.90\times10^{32}$ & $8.93\times10^{22}$ \\ 
$(1,13)\rightarrow(0,11)$ & 1.6989 & $9.37\times10^{0}$ & $2.45\times10^{25}$ & $1.66\times10^{29}$ & $3.55\times10^{31}$ & $3.58\times10^{32}$ & $1.11\times10^{23}$ \\ 
$(1,12)\rightarrow(0,10)$ & 1.7019 & $3.36\times10^{1}$ & $4.43\times10^{26}$ & $1.54\times10^{30}$ & $2.01\times10^{32}$ & $9.96\times10^{32}$ & $3.40\times10^{23}$ \\ 
$(2,20)\rightarrow(0,20)$ & 1.7022 & $4.86\times10^{-2}$ & $6.29\times10^{16}$ & $1.64\times10^{23}$ & $3.60\times10^{27}$ & $1.47\times10^{30}$ & $4.57\times10^{20}$ \\ 
$(1,14)\rightarrow(0,12)$ & 1.7048 & $1.58\times10^{0}$ & $8.06\times10^{23}$ & $1.07\times10^{28}$ & $3.79\times10^{30}$ & $8.03\times10^{31}$ & $2.51\times10^{22}$ \\ 
$(1,11)\rightarrow(0,9)$ & 1.7134 & $9.49\times10^{1}$ & $6.07\times10^{27}$ & $1.10\times10^{31}$ & $8.47\times10^{32}$ & $2.21\times10^{33}$ & $7.61\times10^{23}$ \\ 
$(1,15)\rightarrow(0,13)$ & 1.7203 & $2.19\times10^{-2}$ & $2.19\times10^{21}$ & $5.67\times10^{25}$ & $3.35\times10^{28}$ & $1.12\times10^{30}$ & $3.52\times10^{20}$ \\ 
$(1,10)\rightarrow(0,8)$ & 1.7331 & $2.25\times10^{2}$ & $6.57\times10^{28}$ & $6.38\times10^{31}$ & $2.43\times10^{33}$ & $4.32\times10^{33}$ & $1.73\times10^{24}$ \\ 
$(1,16)\rightarrow(0,14)$ & 1.7464 & $0.23\times10^{0}$ & $4.68\times10^{21}$ & $2.34\times10^{26}$ & $2.28\times10^{29}$ & $1.07\times10^{31}$ & $3.42\times10^{21}$ \\ 
$(2,8)\rightarrow(0,10)$ & 1.7552 & $1.92\times10^{0}$ & $3.29\times10^{25}$ & $1.02\times10^{29}$ & $1.23\times10^{31}$ & $1.27\times10^{32}$ & $4.08\times10^{22}$ \\ 
$(1,9)\rightarrow(0,7)$ & 1.7610 & $4.66\times10^{2}$ & $5.76\times10^{29}$ & $3.09\times10^{32}$ & $5.01\times10^{33}$ & $7.29\times10^{33}$ & $2.97\times10^{24}$ \\ 
$(1,17)\rightarrow(0,15)$ & 1.7847 & $0.65\times10^{0}$ & $2.76\times10^{21}$ & $2.60\times10^{26}$ & $4.15\times10^{29}$ & $2.62\times10^{31}$ & $8.55\times10^{21}$ \\ 
$(1,8)\rightarrow(0,6)$ & 1.7974 & $8.55\times10^{2}$ & $4.05\times10^{30}$ & $1.16\times10^{33}$ & $8.32\times10^{33}$ & $1.13\times10^{34}$ & $4.72\times10^{24}$ \\ 
$(2,12)\rightarrow(1,10)$ & 1.8248 & $2.55\times10^{0}$ & $1.85\times10^{23}$ & $5.41\times10^{27}$ & $3.51\times10^{30}$ & $1.02\times10^{32}$ & $3.39\times10^{22}$ \\ 
$(2,13)\rightarrow(1,11)$ & 1.8259 & $0.40\times10^{0}$ & $6.58\times10^{21}$ & $3.55\times10^{26}$ & $3.68\times10^{29}$ & $1.77\times10^{31}$ & $5.93\times10^{21}$ \\ 
$(2,9)\rightarrow(0,11)$ & 1.8302 & $0.75\times10^{0}$ & $3.60\times10^{24}$ & $1.87\times10^{28}$ & $3.30\times10^{30}$ & $4.74\times10^{31}$ & $1.59\times10^{22}$ \\ 
$(2,11)\rightarrow(1,9)$ & 1.8333 & $8.88\times10^{0}$ & $2.76\times10^{24}$ & $4.44\times10^{28}$ & $1.82\times10^{31}$ & $2.81\times10^{32}$ & $9.42\times10^{22}$ \\ 
$(2,14)\rightarrow(1,12)$ & 1.8371 & $1.74\times10^{-4}$ & $6.35\times10^{17}$ & $6.33\times10^{22}$ & $1.06\times10^{26}$ & $7.16\times10^{27}$ & $2.41\times10^{18}$ \\ 
$(1,18)\rightarrow(0,16)$ & 1.8372 & $0.89\times10^{0}$ & $8.63\times10^{20}$ & $1.48\times10^{26}$ & $3.77\times10^{29}$ & $3.20\times10^{31}$ & $1.08\times10^{22}$ \\ 
\hline
\end{tabular}
\end{table*}

\begin{table*}
\contcaption{}
\begin{tabular}{lccccccc}
\hline
$(v_{\rm i},J_{\rm i})\rightarrow(v_{\rm f},J_{\rm f})$ &
$\lambda[\mu{\rm m}]$ & $\tau$ & $L_{650}$ & $L_{1000}$ & $L_{1500}$ &
$L_{\tau_{\rm c}=1}$ & $l_{\tau_{\rm c}=1}$ \\
\hline
$(1,7)\rightarrow(0,5)$ & 1.8426 & $1.39\times10^{3}$ & $2.26\times10^{31}$ & $3.04\times10^{33}$ & $1.26\times10^{34}$ & $1.55\times10^{34}$ & $8.12\times10^{24}$ \\ 
$(2,10)\rightarrow(1,8)$ & 1.8511 & $2.34\times10^{1}$ & $2.96\times10^{25}$ & $2.67\times10^{29}$ & $7.08\times10^{31}$ & $6.21\times10^{32}$ & $2.31\times10^{23}$ \\ 
$(2,15)\rightarrow(1,13)$ & 1.8595 & $0.13\times10^{0}$ & $1.03\times10^{20}$ & $1.90\times10^{25}$ & $5.11\times10^{28}$ & $4.80\times10^{30}$ & $1.64\times10^{21}$ \\ 
$(2,9)\rightarrow(1,7)$ & 1.8780 & $5.15\times10^{1}$ & $2.45\times10^{26}$ & $1.28\times10^{30}$ & $2.25\times10^{32}$ & $1.18\times10^{33}$ & $4.44\times10^{23}$ \\ 
$(2,16)\rightarrow(1,14)$ & 1.8944 & $0.31\times10^{0}$ & $5.73\times10^{19}$ & $1.91\times10^{25}$ & $8.24\times10^{28}$ & $1.06\times10^{31}$ & $3.69\times10^{21}$ \\ 
$(1,6)\rightarrow(0,4)$ & 1.8971 & $1.99\times10^{3}$ & $9.77\times10^{31}$ & $5.83\times10^{33}$ & $1.68\times10^{34}$ & $2.01\times10^{34}$ & $1.09\times10^{25}$ \\ 
$(1,19)\rightarrow(0,17)$ & 1.9070 & $0.97\times10^{0}$ & $2.37\times10^{20}$ & $7.08\times10^{25}$ & $2.79\times10^{29}$ & $3.16\times10^{31}$ & $1.10\times10^{22}$ \\ 
$(2,10)\rightarrow(0,12)$ & 1.9071 & $0.27\times10^{0}$ & $3.32\times10^{23}$ & $3.00\times10^{27}$ & $7.96\times10^{29}$ & $1.55\times10^{31}$ & $5.42\times10^{21}$ \\ 
$(2,8)\rightarrow(1,6)$ & 1.9142 & $9.77\times10^{1}$ & $1.61\times10^{27}$ & $5.00\times10^{30}$ & $5.94\times10^{32}$ & $2.02\times10^{33}$ & $7.78\times10^{23}$ \\ 
$(2,17)\rightarrow(1,15)$ & 1.9439 & $0.41\times10^{0}$ & $1.85\times10^{19}$ & $1.10\times10^{25}$ & $7.45\times10^{28}$ & $1.31\times10^{31}$ & $4.65\times10^{21}$ \\ 
$(2,7)\rightarrow(1,5)$ & 1.9600 & $1.62\times10^{2}$ & $8.29\times10^{27}$ & $1.61\times10^{31}$ & $1.27\times10^{33}$ & $3.06\times10^{33}$ & $1.20\times10^{24}$ \\ 
$(1,5)\rightarrow(0,3)$ & 1.9617 & $2.51\times10^{3}$ & $3.21\times10^{32}$ & $9.35\times10^{33}$ & $2.20\times10^{34}$ & $2.55\times10^{34}$ & $1.43\times10^{25}$ \\ 
$(2,11)\rightarrow(0,13)$ & 1.9845 & $8.94\times10^{-2}$ & $2.65\times10^{22}$ & $4.27\times10^{26}$ & $1.75\times10^{29}$ & $4.59\times10^{30}$ & $1.67\times10^{21}$ \\ 
$(1,20)\rightarrow(0,18)$ & 1.9985 & $0.95\times10^{0}$ & $6.70\times10^{19}$ & $3.30\times10^{25}$ & $1.93\times10^{29}$ & $2.87\times10^{31}$ & $1.05\times10^{22}$ \\ 
$(2,18)\rightarrow(1,16)$ & 2.0108 & $0.44\times10^{0}$ & $5.19\times10^{18}$ & $5.29\times10^{24}$ & $5.51\times10^{28}$ & $1.30\times10^{31}$ & $4.79\times10^{21}$ \\ 
$(2,6)\rightarrow(1,4)$ & 2.0160 & $2.35\times10^{2}$ & $3.34\times10^{28}$ & $4.23\times10^{31}$ & $2.21\times10^{33}$ & $4.37\times10^{33}$ & $2.04\times10^{24}$ \\ 
$(1,4)\rightarrow(0,2)$ & 2.0372 & $2.70\times10^{3}$ & $7.62\times10^{32}$ & $1.28\times10^{34}$ & $2.68\times10^{34}$ & $2.95\times10^{34}$ & $1.71\times10^{25}$ \\ 
$(2,12)\rightarrow(0,14)$ & 2.0602 & $2.80\times10^{-2}$ & $1.88\times10^{21}$ & $5.50\times10^{25}$ & $3.57\times10^{28}$ & $1.27\times10^{30}$ & $4.78\times10^{20}$ \\ 
$(2,5)\rightarrow(1,3)$ & 2.0829 & $2.98\times10^{2}$ & $1.03\times10^{29}$ & $8.95\times10^{31}$ & $3.17\times10^{33}$ & $5.33\times10^{33}$ & $2.57\times10^{24}$ \\ 
$(2,19)\rightarrow(1,17)$ & 2.0994 & $0.43\times10^{0}$ & $1.45\times10^{18}$ & $2.43\times10^{24}$ & $3.76\times10^{28}$ & $1.17\times10^{31}$ & $4.51\times10^{21}$ \\ 
$(1,3)\rightarrow(0,1)$ & 2.1249 & $2.40\times10^{3}$ & $1.28\times10^{33}$ & $1.62\times10^{34}$ & $3.05\times10^{34}$ & $3.40\times10^{34}$ & $2.06\times10^{25}$ \\ 
$(2,13)\rightarrow(0,15)$ & 2.1320 & $8.21\times10^{-3}$ & $1.21\times10^{20}$ & $6.52\times10^{24}$ & $6.76\times10^{27}$ & $3.31\times10^{29}$ & $1.29\times10^{20}$ \\ 
$(2,4)\rightarrow(1,2)$ & 2.1617 & $3.25\times10^{2}$ & $2.36\times10^{29}$ & $1.49\times10^{32}$ & $3.97\times10^{33}$ & $6.30\times10^{33}$ & $3.15\times10^{24}$ \\ 
$(2,14)\rightarrow(0,16)$ & 2.1968 & $2.27\times10^{-3}$ & $7.26\times10^{18}$ & $7.25\times10^{23}$ & $1.21\times10^{27}$ & $8.20\times10^{28}$ & $3.30\times10^{19}$ \\ 
$(2,20)\rightarrow(1,18)$ & 2.2155 & $0.41\times10^{0}$ & $4.45\times10^{17}$ & $1.16\times10^{24}$ & $2.55\times10^{28}$ & $1.02\times10^{31}$ & $4.13\times10^{21}$ \\ 
$(1,2)\rightarrow(0,0)$ & 2.2261 & $1.54\times10^{3}$ & $1.33\times10^{33}$ & $1.67\times10^{34}$ & $3.19\times10^{34}$ & $3.53\times10^{34}$ & $2.24\times10^{25}$ \\ 
$(2,15)\rightarrow(0,17)$ & 2.2514 & $6.02\times10^{-4}$ & $4.22\times10^{17}$ & $7.76\times10^{22}$ & $2.09\times10^{26}$ & $1.96\times10^{28}$ & $8.10\times10^{18}$ \\ 
$(2,3)\rightarrow(1,1)$ & 2.2537 & $2.91\times10^{2}$ & $3.81\times10^{29}$ & $1.86\times10^{32}$ & $4.20\times10^{33}$ & $6.44\times10^{33}$ & $3.36\times10^{24}$ \\ 
$(2,16)\rightarrow(0,18)$ & 2.2922 & $1.55\times10^{-4}$ & $2.48\times10^{16}$ & $8.29\times10^{21}$ & $3.57\times10^{25}$ & $4.64\times10^{27}$ & $1.95\times10^{18}$ \\ 
$(2,17)\rightarrow(0,19)$ & 2.3159 & $4.12\times10^{-5}$ & $1.60\times10^{15}$ & $9.48\times10^{20}$ & $6.42\times10^{24}$ & $1.15\times10^{27}$ & $4.86\times10^{17}$ \\ 
$(2,18)\rightarrow(0,20)$ & 2.3196 & $1.26\times10^{-5}$ & $1.29\times10^{14}$ & $1.32\times10^{20}$ & $1.37\times10^{24}$ & $3.31\times10^{26}$ & $1.40\times10^{17}$ \\ 
$(2,2)\rightarrow(1,0)$ & 2.3603 & $1.88\times10^{2}$ & $3.78\times10^{29}$ & $1.52\times10^{32}$ & $3.38\times10^{33}$ & $5.33\times10^{33}$ & $2.91\times10^{24}$ \\ 
$(1,1)\rightarrow(0,1)$ & 2.4094 & $1.82\times10^{3}$ & $2.04\times10^{33}$ & $2.21\times10^{34}$ & $3.90\times10^{34}$ & $4.26\times10^{34}$ & $2.92\times10^{25}$ \\ 
$(1,2)\rightarrow(0,2)$ & 2.4165 & $1.85\times10^{3}$ & $1.49\times10^{33}$ & $2.04\times10^{34}$ & $3.64\times10^{34}$ & $3.98\times10^{34}$ & $2.74\times10^{25}$ \\ 
$(1,3)\rightarrow(0,3)$ & 2.4273 & $1.92\times10^{3}$ & $9.28\times10^{32}$ & $1.77\times10^{34}$ & $3.41\times10^{34}$ & $3.76\times10^{34}$ & $2.60\times10^{25}$ \\ 
$(1,4)\rightarrow(0,4)$ & 2.4416 & $1.78\times10^{3}$ & $4.34\times10^{32}$ & $1.36\times10^{34}$ & $3.03\times10^{34}$ & $3.27\times10^{34}$ & $2.27\times10^{25}$ \\ 
$(1,5)\rightarrow(0,5)$ & 2.4597 & $1.49\times10^{3}$ & $1.56\times10^{32}$ & $9.01\times10^{33}$ & $2.37\times10^{34}$ & $2.72\times10^{34}$ & $1.91\times10^{25}$ \\ 
$(1,6)\rightarrow(0,6)$ & 2.4815 & $1.13\times10^{3}$ & $4.37\times10^{31}$ & $4.65\times10^{33}$ & $1.81\times10^{34}$ & $2.08\times10^{34}$ & $1.20\times10^{25}$ \\ 
$(1,7)\rightarrow(0,7)$ & 2.5071 & $7.88\times10^{2}$ & $9.81\times10^{30}$ & $1.81\times10^{33}$ & $1.19\times10^{34}$ & $1.46\times10^{34}$ & $8.49\times10^{24}$ \\ 
$(1,8)\rightarrow(0,8)$ & 2.5366 & $5.12\times10^{2}$ & $1.81\times10^{30}$ & $5.59\times10^{32}$ & $7.21\times10^{33}$ & $9.57\times10^{33}$ & $5.62\times10^{24}$ \\ 
$(2,1)\rightarrow(1,1)$ & 2.5547 & $2.26\times10^{2}$ & $5.72\times10^{29}$ & $2.02\times10^{32}$ & $3.92\times10^{33}$ & $5.77\times10^{33}$ & $3.41\times10^{24}$ \\ 
$(2,2)\rightarrow(1,2)$ & 2.5633 & $2.34\times10^{2}$ & $4.26\times10^{29}$ & $1.71\times10^{32}$ & $3.74\times10^{33}$ & $5.77\times10^{33}$ & $3.42\times10^{24}$ \\ 
$(1,9)\rightarrow(0,9)$ & 2.5701 & $3.13\times10^{2}$ & $2.81\times10^{29}$ & $1.51\times10^{32}$ & $3.84\times10^{33}$ & $5.84\times10^{33}$ & $3.47\times10^{24}$ \\ 
$(2,3)\rightarrow(1,3)$ & 2.5763 & $2.46\times10^{2}$ & $2.76\times10^{29}$ & $1.35\times10^{32}$ & $3.41\times10^{33}$ & $5.40\times10^{33}$ & $3.22\times10^{24}$ \\ 
$(2,4)\rightarrow(1,4)$ & 2.5937 & $2.32\times10^{2}$ & $1.37\times10^{29}$ & $8.68\times10^{31}$ & $2.80\times10^{33}$ & $4.59\times10^{33}$ & $2.76\times10^{24}$ \\ 
$(1,10)\rightarrow(0,10)$ & 2.6077 & $1.80\times10^{2}$ & $3.72\times10^{28}$ & $3.60\times10^{31}$ & $1.71\times10^{33}$ & $3.49\times10^{33}$ & $1.83\times10^{24}$ \\ 
$(2,5)\rightarrow(1,5)$ & 2.6156 & $1.98\times10^{2}$ & $5.31\times10^{28}$ & $4.62\times10^{31}$ & $1.99\times10^{33}$ & $3.72\times10^{33}$ & $1.95\times10^{24}$ \\ 
$(1,0)\rightarrow(0,2)$ & 2.6297 & $1.32\times10^{3}$ & $1.63\times10^{33}$ & $2.18\times10^{34}$ & $3.79\times10^{34}$ & $4.11\times10^{34}$ & $3.08\times10^{25}$ \\ 
$(2,6)\rightarrow(1,6)$ & 2.6421 & $1.54\times10^{2}$ & $1.64\times10^{28}$ & $2.07\times10^{31}$ & $1.24\times10^{33}$ & $2.90\times10^{33}$ & $1.54\times10^{24}$ \\ 
$(1,11)\rightarrow(0,11)$ & 2.6495 & $9.92\times10^{1}$ & $4.37\times10^{27}$ & $7.92\times10^{30}$ & $6.37\times10^{32}$ & $1.91\times10^{33}$ & $1.02\times10^{24}$ \\ 
$(2,7)\rightarrow(1,7)$ & 2.6735 & $1.11\times10^{2}$ & $4.09\times10^{27}$ & $7.93\times10^{30}$ & $6.68\times10^{32}$ & $1.99\times10^{33}$ & $1.07\times10^{24}$ \\ 
$(1,12)\rightarrow(0,12)$ & 2.6955 & $5.26\times10^{1}$ & $4.64\times10^{26}$ & $1.61\times10^{30}$ & $2.11\times10^{32}$ & $9.93\times10^{32}$ & $5.37\times10^{23}$ \\ 
$(2,8)\rightarrow(1,8)$ & 2.7097 & $7.41\times10^{1}$ & $8.51\times10^{26}$ & $2.65\times10^{30}$ & $3.18\times10^{32}$ & $1.28\times10^{33}$ & $6.94\times10^{23}$ \\ 
$(1,13)\rightarrow(0,13)$ & 2.7459 & $2.71\times10^{1}$ & $4.60\times10^{25}$ & $3.11\times10^{29}$ & $6.67\times10^{31}$ & $5.20\times10^{32}$ & $2.87\times10^{23}$ \\ 
$(2,9)\rightarrow(1,9)$ & 2.7510 & $4.68\times10^{1}$ & $1.50\times10^{26}$ & $7.81\times10^{29}$ & $1.38\times10^{32}$ & $8.02\times10^{32}$ & $4.43\times10^{23}$ \\ 
$(2,0)\rightarrow(1,2)$ & 2.7891 & $1.69\times10^{2}$ & $4.45\times10^{29}$ & $1.47\times10^{32}$ & $2.99\times10^{33}$ & $4.57\times10^{33}$ & $2.95\times10^{24}$ \\ 
$(2,10)\rightarrow(1,10)$ & 2.7976 & $2.81\times10^{1}$ & $2.31\times10^{25}$ & $2.08\times10^{29}$ & $5.52\times10^{31}$ & $4.79\times10^{32}$ & $2.69\times10^{23}$ \\ 
$(1,14)\rightarrow(0,14)$ & 2.8008 & $1.36\times10^{1}$ & $4.37\times10^{24}$ & $5.79\times10^{28}$ & $2.06\times10^{31}$ & $2.71\times10^{32}$ & $1.39\times10^{23}$ \\ 
$(1,1)\rightarrow(0,3)$ & 2.8057 & $1.87\times10^{3}$ & $1.77\times10^{33}$ & $2.49\times10^{34}$ & $4.22\times10^{34}$ & $4.55\times10^{34}$ & $3.64\times10^{25}$ \\ 
$(2,11)\rightarrow(1,11)$ & 2.8498 & $1.60\times10^{1}$ & $3.14\times10^{24}$ & $5.04\times10^{28}$ & $2.07\times10^{31}$ & $2.76\times10^{32}$ & $1.44\times10^{23}$ \\ 
$(1,15)\rightarrow(0,15)$ & 2.8601 & $6.79\times10^{0}$ & $4.14\times10^{23}$ & $1.07\times10^{28}$ & $6.34\times10^{30}$ & $1.40\times10^{32}$ & $7.35\times10^{22}$ \\ 
$(2,12)\rightarrow(1,12)$ & 2.9077 & $8.82\times10^{0}$ & $3.90\times10^{23}$ & $1.14\times10^{28}$ & $7.39\times10^{30}$ & $1.62\times10^{32}$ & $8.63\times10^{22}$ \\ 
$(1,16)\rightarrow(0,16)$ & 2.9240 & $3.39\times10^{0}$ & $4.03\times10^{22}$ & $2.01\times10^{27}$ & $1.97\times10^{30}$ & $7.09\times10^{31}$ & $3.80\times10^{22}$ \\ 
$(2,13)\rightarrow(1,13)$ & 2.9716 & $4.73\times10^{0}$ & $4.56\times10^{22}$ & $2.46\times10^{27}$ & $2.55\times10^{30}$ & $8.82\times10^{31}$ & $4.80\times10^{22}$ \\ 
\hline
\end{tabular}
\end{table*}

\begin{table*}
\contcaption{}
\begin{tabular}{lccccccc}
\hline
$(v_{\rm i},J_{\rm i})\rightarrow(v_{\rm f},J_{\rm f})$ &
$\lambda[\mu{\rm m}]$ & $\tau$ & $L_{650}$ & $L_{1000}$ & $L_{1500}$ &
$L_{\tau_{\rm c}=1}$ & $l_{\tau_{\rm c}=1}$ \\
\hline
$(2,1)\rightarrow(1,3)$ & 2.9774 & $2.44\times10^{2}$ & $4.94\times10^{29}$ & $1.74\times10^{32}$ & $3.51\times10^{33}$ & $5.26\times10^{33}$ & $3.62\times10^{24}$ \\ 
$(1,17)\rightarrow(0,17)$ & 2.9923 & $1.70\times10^{0}$ & $4.16\times10^{21}$ & $3.92\times10^{26}$ & $6.26\times10^{29}$ & $3.57\times10^{31}$ & $1.96\times10^{22}$ \\ 
$(1,2)\rightarrow(0,4)$ & 3.0073 & $1.92\times10^{3}$ & $1.16\times10^{33}$ & $2.24\times10^{34}$ & $3.94\times10^{34}$ & $4.24\times10^{34}$ & $3.64\times10^{25}$ \\ 
$(2,14)\rightarrow(1,14)$ & 3.0417 & $2.50\times10^{0}$ & $5.17\times10^{21}$ & $5.16\times10^{26}$ & $8.61\times10^{29}$ & $4.72\times10^{31}$ & $2.63\times10^{22}$ \\ 
$(1,18)\rightarrow(0,18)$ & 3.0651 & $0.88\times10^{0}$ & $4.78\times10^{20}$ & $8.20\times10^{25}$ & $2.09\times10^{29}$ & $1.78\times10^{31}$ & $10.00\times10^{21}$ \\ 
$(2,15)\rightarrow(1,15)$ & 3.1184 & $1.30\times10^{0}$ & $5.80\times10^{20}$ & $1.07\times10^{26}$ & $2.87\times10^{29}$ & $2.43\times10^{31}$ & $1.39\times10^{22}$ \\ 
$(1,19)\rightarrow(0,19)$ & 3.1422 & $0.47\times10^{0}$ & $6.28\times10^{19}$ & $1.88\times10^{25}$ & $7.40\times10^{28}$ & $8.86\times10^{30}$ & $5.10\times10^{21}$ \\ 
$(2,2)\rightarrow(1,4)$ & 3.1941 & $2.56\times10^{2}$ & $3.35\times10^{29}$ & $1.35\times10^{32}$ & $3.07\times10^{33}$ & $4.63\times10^{33}$ & $3.42\times10^{24}$ \\ 
$(2,16)\rightarrow(1,16)$ & 3.2018 & $0.68\times10^{0}$ & $6.72\times10^{19}$ & $2.25\times10^{25}$ & $9.66\times10^{28}$ & $1.20\times10^{31}$ & $7.05\times10^{21}$ \\ 
$(1,20)\rightarrow(0,20)$ & 3.2232 & $0.26\times10^{0}$ & $9.88\times10^{18}$ & $4.86\times10^{24}$ & $2.84\times10^{28}$ & $4.55\times10^{30}$ & $2.69\times10^{21}$ \\ 
$(1,3)\rightarrow(0,5)$ & 3.2385 & $1.65\times10^{3}$ & $5.24\times10^{32}$ & $1.61\times10^{34}$ & $3.22\times10^{34}$ & $3.37\times10^{34}$ & $3.11\times10^{25}$ \\ 
$(2,17)\rightarrow(1,17)$ & 3.2923 & $0.36\times10^{0}$ & $8.30\times10^{18}$ & $4.92\times10^{24}$ & $3.33\times10^{28}$ & $5.88\times10^{30}$ & $3.54\times10^{21}$ \\ 
$(2,18)\rightarrow(1,18)$ & 3.3899 & $0.19\times10^{0}$ & $1.12\times10^{18}$ & $1.14\times10^{24}$ & $1.19\times10^{28}$ & $2.86\times10^{30}$ & $1.78\times10^{21}$ \\ 
$(2,3)\rightarrow(1,5)$ & 3.4438 & $2.25\times10^{2}$ & $1.59\times10^{29}$ & $7.79\times10^{31}$ & $2.18\times10^{33}$ & $3.57\times10^{33}$ & $2.47\times10^{24}$ \\ 
$(2,19)\rightarrow(1,19)$ & 3.4948 & $0.11\times10^{0}$ & $1.75\times10^{17}$ & $2.93\times10^{23}$ & $4.53\times10^{27}$ & $1.44\times10^{30}$ & $9.22\times10^{20}$ \\ 
$(1,4)\rightarrow(0,6)$ & 3.5034 & $1.22\times10^{3}$ & $1.71\times10^{32}$ & $8.63\times10^{33}$ & $2.22\times10^{34}$ & $2.37\times10^{34}$ & $1.92\times10^{25}$ \\ 
$(2,20)\rightarrow(1,20)$ & 3.6070 & $6.16\times10^{-2}$ & $3.23\times10^{16}$ & $8.39\times10^{22}$ & $1.85\times10^{27}$ & $7.53\times10^{29}$ & $4.97\times10^{20}$ \\ 
$(2,4)\rightarrow(1,6)$ & 3.7313 & $1.72\times10^{2}$ & $5.56\times10^{28}$ & $3.52\times10^{31}$ & $1.26\times10^{33}$ & $2.37\times10^{33}$ & $1.78\times10^{24}$ \\ 
$(1,5)\rightarrow(0,7)$ & 3.8063 & $8.07\times10^{2}$ & $4.18\times10^{31}$ & $3.29\times10^{33}$ & $1.33\times10^{34}$ & $1.49\times10^{34}$ & $1.31\times10^{25}$ \\ 
$(2,5)\rightarrow(1,7)$ & 4.0622 & $1.18\times10^{2}$ & $1.48\times10^{28}$ & $1.29\times10^{31}$ & $6.06\times10^{32}$ & $1.45\times10^{33}$ & $1.19\times10^{24}$ \\ 
$(1,6)\rightarrow(0,8)$ & 4.1517 & $4.84\times10^{2}$ & $7.87\times10^{30}$ & $9.33\times10^{32}$ & $6.68\times10^{33}$ & $7.78\times10^{33}$ & $7.48\times10^{24}$ \\ 
$(0,15)\rightarrow(0,13)$ & 4.3167 & $1.04\times10^{3}$ & $5.78\times10^{27}$ & $1.71\times10^{31}$ & $1.63\times10^{33}$ & $3.08\times10^{33}$ & $3.08\times10^{24}$ \\ 
$(0,16)\rightarrow(0,14)$ & 4.3364 & $6.68\times10^{2}$ & $6.49\times10^{26}$ & $3.93\times10^{30}$ & $7.34\times10^{32}$ & $1.86\times10^{33}$ & $1.62\times10^{24}$ \\ 
$(0,14)\rightarrow(0,12)$ & 4.3568 & $1.59\times10^{3}$ & $5.08\times10^{28}$ & $7.30\times10^{31}$ & $3.22\times10^{33}$ & $4.75\times10^{33}$ & $4.79\times10^{24}$ \\ 
$(0,17)\rightarrow(0,15)$ & 4.4219 & $4.35\times10^{2}$ & $7.52\times10^{25}$ & $9.09\times10^{29}$ & $3.00\times10^{32}$ & $1.18\times10^{33}$ & $1.05\times10^{24}$ \\ 
$(2,6)\rightarrow(1,8)$ & 4.4422 & $7.29\times10^{1}$ & $3.05\times10^{27}$ & $3.86\times10^{30}$ & $2.40\times10^{32}$ & $7.47\times10^{32}$ & $6.66\times10^{23}$ \\ 
$(0,13)\rightarrow(0,11)$ & 4.4558 & $2.38\times10^{3}$ & $4.26\times10^{29}$ & $2.96\times10^{32}$ & $5.79\times10^{33}$ & $7.21\times10^{33}$ & $7.44\times10^{24}$ \\ 
$(1,7)\rightarrow(0,9)$ & 4.5431 & $2.63\times10^{2}$ & $1.15\times10^{30}$ & $2.15\times10^{32}$ & $2.80\times10^{33}$ & $3.74\times10^{33}$ & $3.93\times10^{24}$ \\ 
$(0,18)\rightarrow(0,16)$ & 4.5859 & $2.89\times10^{2}$ & $9.34\times10^{24}$ & $2.18\times10^{29}$ & $1.19\times10^{32}$ & $7.07\times10^{32}$ & $6.51\times10^{23}$ \\ 
$(1,15)\rightarrow(1,13)$ & 4.6058 & $1.07\times10^{2}$ & $2.65\times10^{24}$ & $6.86\times10^{28}$ & $4.05\times10^{31}$ & $3.80\times10^{32}$ & $3.52\times10^{23}$ \\ 
$(0,12)\rightarrow(0,10)$ & 4.6173 & $3.43\times10^{3}$ & $3.27\times10^{30}$ & $1.11\times10^{33}$ & $9.48\times10^{33}$ & $1.08\times10^{34}$ & $1.16\times10^{25}$ \\ 
$(1,14)\rightarrow(1,12)$ & 4.6386 & $1.58\times10^{2}$ & $1.97\times10^{25}$ & $2.62\times10^{29}$ & $9.29\times10^{31}$ & $5.62\times10^{32}$ & $5.23\times10^{23}$ \\ 
$(1,16)\rightarrow(1,14)$ & 4.6391 & $7.20\times10^{1}$ & $3.55\times10^{23}$ & $1.77\times10^{28}$ & $1.73\times10^{31}$ & $2.43\times10^{32}$ & $2.27\times10^{23}$ \\ 
$(1,13)\rightarrow(1,11)$ & 4.7356 & $2.28\times10^{2}$ & $1.40\times10^{26}$ & $9.44\times10^{29}$ & $2.02\times10^{32}$ & $8.39\times10^{32}$ & $7.98\times10^{23}$ \\ 
$(1,17)\rightarrow(1,15)$ & 4.7462 & $4.87\times10^{1}$ & $4.86\times10^{22}$ & $4.58\times10^{27}$ & $7.31\times10^{30}$ & $1.61\times10^{32}$ & $1.53\times10^{23}$ \\ 
$(0,11)\rightarrow(0,9)$ & 4.8495 & $4.68\times10^{3}$ & $2.20\times10^{31}$ & $3.64\times10^{33}$ & $1.49\times10^{34}$ & $1.65\times10^{34}$ & $1.86\times10^{25}$ \\ 
$(0,19)\rightarrow(0,17)$ & 4.8513 & $2.02\times10^{2}$ & $1.30\times10^{24}$ & $5.58\times10^{28}$ & $4.86\times10^{31}$ & $4.45\times10^{32}$ & $4.33\times10^{23}$ \\ 
$(2,7)\rightarrow(1,9)$ & 4.8764 & $4.10\times10^{1}$ & $4.95\times10^{26}$ & $9.59\times10^{29}$ & $8.15\times10^{31}$ & $3.54\times10^{32}$ & $3.46\times10^{23}$ \\ 
$(1,12)\rightarrow(1,10)$ & 4.8999 & $3.17\times10^{2}$ & $9.11\times10^{26}$ & $3.17\times10^{30}$ & $4.07\times10^{32}$ & $1.13\times10^{33}$ & $1.11\times10^{24}$ \\ 
$(1,18)\rightarrow(1,16)$ & 4.9429 & $3.40\times10^{1}$ & $7.16\times10^{21}$ & $1.23\times10^{27}$ & $3.13\times10^{30}$ & $1.03\times10^{32}$ & $9.32\times10^{22}$ \\ 
$(2,15)\rightarrow(2,13)$ & 4.9684 & $1.31\times10^{1}$ & $2.31\times10^{21}$ & $4.24\times10^{26}$ & $1.14\times10^{30}$ & $5.49\times10^{31}$ & $4.99\times10^{22}$ \\ 
$(1,8)\rightarrow(0,10)$ & 4.9824 & $1.31\times10^{2}$ & $1.35\times10^{29}$ & $4.17\times10^{31}$ & $9.06\times10^{32}$ & $1.52\times10^{33}$ & $1.52\times10^{24}$ \\ 
$(2,14)\rightarrow(2,12)$ & 4.9894 & $1.86\times10^{1}$ & $1.44\times10^{22}$ & $1.44\times10^{27}$ & $2.40\times10^{30}$ & $7.54\times10^{31}$ & $6.89\times10^{22}$ \\ 
$(2,16)\rightarrow(2,14)$ & 5.0221 & $9.20\times10^{0}$ & $3.67\times10^{20}$ & $1.23\times10^{26}$ & $5.27\times10^{29}$ & $3.84\times10^{31}$ & $3.53\times10^{22}$ \\ 
$(2,13)\rightarrow(2,11)$ & 5.0817 & $2.57\times10^{1}$ & $8.60\times10^{22}$ & $4.64\times10^{27}$ & $4.80\times10^{30}$ & $1.02\times10^{32}$ & $9.49\times10^{22}$ \\ 
$(1,11)\rightarrow(1,9)$ & 5.1400 & $4.19\times10^{2}$ & $5.27\times10^{27}$ & $9.57\times10^{30}$ & $7.26\times10^{32}$ & $1.52\times10^{33}$ & $1.57\times10^{24}$ \\ 
$(2,17)\rightarrow(2,15)$ & 5.1610 & $6.57\times10^{0}$ & $6.04\times10^{19}$ & $3.59\times10^{25}$ & $2.43\times10^{29}$ & $2.64\times10^{31}$ & $2.49\times10^{22}$ \\ 
$(0,10)\rightarrow(0,8)$ & 5.1673 & $5.98\times10^{3}$ & $1.26\times10^{32}$ & $9.17\times10^{33}$ & $2.09\times10^{34}$ & $2.29\times10^{34}$ & $3.38\times10^{25}$ \\ 
$(2,12)\rightarrow(2,10)$ & 5.2478 & $3.45\times10^{1}$ & $4.75\times10^{23}$ & $1.39\times10^{28}$ & $9.01\times10^{30}$ & $1.32\times10^{32}$ & $1.40\times10^{23}$ \\ 
$(1,19)\rightarrow(1,17)$ & 5.2578 & $2.45\times10^{1}$ & $1.16\times10^{21}$ & $3.46\times10^{26}$ & $1.36\times10^{30}$ & $6.60\times10^{31}$ & $6.36\times10^{22}$ \\ 
$(0,20)\rightarrow(0,18)$ & 5.2594 & $1.51\times10^{2}$ & $2.09\times10^{23}$ & $1.56\times10^{28}$ & $2.09\times10^{31}$ & $2.70\times10^{32}$ & $2.85\times10^{23}$ \\ 
$(2,8)\rightarrow(1,10)$ & 5.3686 & $2.12\times10^{1}$ & $6.51\times10^{25}$ & $2.02\times10^{29}$ & $2.44\times10^{31}$ & $1.56\times10^{32}$ & $1.69\times10^{23}$ \\ 
$(2,18)\rightarrow(2,16)$ & 5.4058 & $4.82\times10^{0}$ & $1.06\times10^{19}$ & $1.08\times10^{25}$ & $1.13\times10^{29}$ & $1.79\times10^{31}$ & $1.77\times10^{22}$ \\ 
$(1,9)\rightarrow(0,11)$ & 5.4679 & $6.01\times10^{1}$ & $1.29\times10^{28}$ & $6.91\times10^{30}$ & $2.32\times10^{32}$ & $5.77\times10^{32}$ & $6.34\times10^{23}$ \\ 
$(1,10)\rightarrow(1,8)$ & 5.4711 & $5.17\times10^{2}$ & $2.60\times10^{28}$ & $2.52\times10^{31}$ & $1.12\times10^{33}$ & $1.90\times10^{33}$ & $2.40\times10^{24}$ \\ 
$(2,11)\rightarrow(2,9)$ & 5.4959 & $4.40\times10^{1}$ & $2.34\times10^{24}$ & $3.77\times10^{28}$ & $1.55\times10^{31}$ & $1.56\times10^{32}$ & $1.73\times10^{23}$ \\ 
$(0,9)\rightarrow(0,7)$ & 5.5942 & $6.98\times10^{3}$ & $5.82\times10^{32}$ & $1.72\times10^{34}$ & $2.80\times10^{34}$ & $2.97\times10^{34}$ & $4.73\times10^{25}$ \\ 
$(1,20)\rightarrow(1,18)$ & 5.7428 & $1.96\times10^{1}$ & $2.22\times10^{20}$ & $1.09\times10^{26}$ & $6.38\times10^{29}$ & $4.37\times10^{31}$ & $4.60\times10^{22}$ \\ 
$(2,19)\rightarrow(2,17)$ & 5.7939 & $3.71\times10^{0}$ & $2.06\times10^{18}$ & $3.45\times10^{24}$ & $5.34\times10^{28}$ & $1.19\times10^{31}$ & $1.27\times10^{22}$ \\ 
$(2,10)\rightarrow(2,8)$ & 5.8420 & $5.26\times10^{1}$ & $9.91\times10^{24}$ & $8.94\times10^{28}$ & $2.37\times10^{31}$ & $1.86\times10^{32}$ & $2.18\times10^{23}$ \\ 
$(1,9)\rightarrow(1,7)$ & 5.9180 & $5.88\times10^{2}$ & $1.05\times10^{29}$ & $5.61\times10^{31}$ & $1.51\times10^{33}$ & $2.17\times10^{33}$ & $2.97\times10^{24}$ \\ 
$(2,9)\rightarrow(1,11)$ & 5.9189 & $1.01\times10^{1}$ & $7.05\times10^{24}$ & $3.67\times10^{28}$ & $6.46\times10^{30}$ & $6.33\times10^{31}$ & $6.86\times10^{22}$ \\ 
$(1,10)\rightarrow(0,12)$ & 5.9915 & $2.55\times10^{1}$ & $1.03\times10^{27}$ & $1.00\times10^{30}$ & $5.15\times10^{31}$ & $1.98\times10^{32}$ & $2.38\times10^{23}$ \\ 
$(0,8)\rightarrow(0,6)$ & 6.1682 & $7.36\times10^{3}$ & $2.09\times10^{33}$ & $2.61\times10^{34}$ & $3.58\times10^{34}$ & $3.52\times10^{34}$ & $6.19\times10^{25}$ \\ 
\hline
\end{tabular}
\end{table*}

\begin{table*}
\contcaption{}
\begin{tabular}{lccccccc}
\hline
$(v_{\rm i},J_{\rm i})\rightarrow(v_{\rm f},J_{\rm f})$ &
$\lambda[\mu{\rm m}]$ & $\tau$ & $L_{650}$ & $L_{1000}$ & $L_{1500}$ &
$L_{\tau_{\rm c}=1}$ & $l_{\tau_{\rm c}=1}$ \\
\hline
$(2,9)\rightarrow(2,7)$ & 6.3119 & $5.83\times10^{1}$ & $3.48\times10^{25}$ & $1.81\times10^{29}$ & $3.19\times10^{31}$ & $1.92\times10^{32}$ & $2.44\times10^{23}$ \\ 
$(2,20)\rightarrow(2,18)$ & 6.3949 & $3.09\times10^{0}$ & $4.59\times10^{17}$ & $1.19\times10^{24}$ & $2.63\times10^{28}$ & $7.87\times10^{30}$ & $9.21\times10^{21}$ \\ 
$(1,8)\rightarrow(1,6)$ & 6.5205 & $6.06\times10^{2}$ & $3.31\times10^{29}$ & $1.02\times10^{32}$ & $1.72\times10^{33}$ & $2.23\times10^{33}$ & $3.36\times10^{24}$ \\ 
$(2,10)\rightarrow(1,12)$ & 6.5207 & $4.46\times10^{0}$ & $6.44\times10^{23}$ & $5.81\times10^{27}$ & $1.54\times10^{30}$ & $2.33\times10^{31}$ & $2.78\times10^{22}$ \\ 
$(1,11)\rightarrow(0,13)$ & 6.5354 & $9.87\times10^{0}$ & $7.05\times10^{25}$ & $1.28\times10^{29}$ & $1.04\times10^{31}$ & $6.46\times10^{31}$ & $7.73\times10^{22}$ \\ 
$(2,8)\rightarrow(2,6)$ & 6.9478 & $5.83\times10^{1}$ & $9.62\times10^{25}$ & $2.99\times10^{29}$ & $3.60\times10^{31}$ & $1.74\times10^{32}$ & $2.43\times10^{23}$ \\ 
$(0,7)\rightarrow(0,5)$ & 6.9523 & $6.86\times10^{3}$ & $5.52\times10^{33}$ & $3.30\times10^{34}$ & $3.87\times10^{34}$ & $3.92\times10^{34}$ & $7.77\times10^{25}$ \\ 
$(1,12)\rightarrow(0,14)$ & 7.0689 & $3.51\times10^{0}$ & $4.23\times10^{24}$ & $1.47\times10^{28}$ & $1.92\times10^{30}$ & $1.95\times10^{31}$ & $2.53\times10^{22}$ \\ 
$(2,11)\rightarrow(1,13)$ & 7.1563 & $1.81\times10^{0}$ & $5.08\times10^{22}$ & $8.17\times10^{26}$ & $3.36\times10^{29}$ & $7.78\times10^{30}$ & $1.02\times10^{22}$ \\ 
$(1,7)\rightarrow(1,5)$ & 7.3448 & $5.49\times10^{2}$ & $7.79\times10^{29}$ & $1.45\times10^{32}$ & $1.66\times10^{33}$ & $2.06\times10^{33}$ & $3.51\times10^{24}$ \\ 
$(1,18)\rightarrow(0,20)$ & 7.3467 & $4.62\times10^{-4}$ & $3.78\times10^{16}$ & $6.48\times10^{21}$ & $1.65\times10^{25}$ & $1.50\times10^{27}$ & $2.01\times10^{18}$ \\ 
$(1,13)\rightarrow(0,15)$ & 7.5460 & $1.13\times10^{0}$ & $2.28\times10^{23}$ & $1.54\times10^{27}$ & $3.30\times10^{29}$ & $5.28\times10^{30}$ & $7.30\times10^{21}$ \\ 
$(2,12)\rightarrow(1,14)$ & 7.7921 & $0.67\times10^{0}$ & $3.51\times10^{21}$ & $1.03\times10^{26}$ & $6.66\times10^{28}$ & $2.29\times10^{30}$ & $3.27\times10^{21}$ \\ 
$(1,17)\rightarrow(0,19)$ & 7.8090 & $3.27\times10^{-3}$ & $9.91\times10^{17}$ & $9.33\times10^{22}$ & $1.49\times10^{26}$ & $9.79\times10^{27}$ & $1.40\times10^{19}$ \\ 
$(2,7)\rightarrow(2,5)$ & 7.8200 & $5.17\times10^{1}$ & $2.01\times10^{26}$ & $3.89\times10^{29}$ & $3.31\times10^{31}$ & $1.39\times10^{32}$ & $2.18\times10^{23}$ \\ 
$(1,14)\rightarrow(0,16)$ & 7.9088 & $0.33\times10^{0}$ & $1.14\times10^{22}$ & $1.51\times10^{26}$ & $5.35\times10^{28}$ & $1.26\times10^{30}$ & $1.83\times10^{21}$ \\ 
$(0,6)\rightarrow(0,4)$ & 8.0563 & $5.44\times10^{3}$ & $9.89\times10^{33}$ & $3.48\times10^{34}$ & $3.72\times10^{34}$ & $3.74\times10^{34}$ & $8.58\times10^{25}$ \\ 
$(1,16)\rightarrow(0,18)$ & 8.0684 & $1.81\times10^{-2}$ & $2.37\times10^{19}$ & $1.18\times10^{24}$ & $1.16\times10^{27}$ & $5.43\times10^{28}$ & $8.02\times10^{19}$ \\ 
$(1,15)\rightarrow(0,17)$ & 8.0977 & $8.27\times10^{-2}$ & $5.32\times10^{20}$ & $1.38\times10^{25}$ & $8.13\times10^{27}$ & $2.72\times10^{29}$ & $4.03\times10^{20}$ \\ 
$(2,18)\rightarrow(1,20)$ & 8.2737 & $8.43\times10^{-5}$ & $6.52\times10^{13}$ & $6.64\times10^{19}$ & $6.91\times10^{23}$ & $1.67\times10^{26}$ & $2.52\times10^{17}$ \\ 
$(2,13)\rightarrow(1,15)$ & 8.3749 & $0.22\times10^{0}$ & $2.20\times10^{20}$ & $1.18\times10^{25}$ & $1.23\times10^{28}$ & $5.98\times10^{29}$ & $9.18\times10^{20}$ \\ 
$(1,6)\rightarrow(1,4)$ & 8.5070 & $4.27\times10^{2}$ & $1.28\times10^{30}$ & $1.52\times10^{32}$ & $1.31\times10^{33}$ & $1.56\times10^{33}$ & $3.08\times10^{24}$ \\ 
$(2,17)\rightarrow(1,19)$ & 8.8069 & $6.51\times10^{-4}$ & $1.61\times10^{15}$ & $9.55\times10^{20}$ & $6.47\times10^{24}$ & $1.15\times10^{27}$ & $1.86\times10^{18}$ \\ 
$(2,14)\rightarrow(1,16)$ & 8.8339 & $6.61\times10^{-2}$ & $1.27\times10^{19}$ & $1.26\times10^{24}$ & $2.11\times10^{27}$ & $1.43\times10^{29}$ & $2.31\times10^{20}$ \\ 
$(2,6)\rightarrow(2,4)$ & 9.0514 & $3.94\times10^{1}$ & $2.97\times10^{26}$ & $3.75\times10^{29}$ & $2.34\times10^{31}$ & $9.27\times10^{31}$ & $1.68\times10^{23}$ \\ 
$(2,16)\rightarrow(1,18)$ & 9.0901 & $3.72\times10^{-3}$ & $3.45\times10^{16}$ & $1.15\times10^{22}$ & $4.96\times10^{25}$ & $6.45\times10^{27}$ & $1.07\times10^{19}$ \\ 
$(2,15)\rightarrow(1,17)$ & 9.0924 & $1.71\times10^{-2}$ & $6.82\times10^{17}$ & $1.25\times10^{23}$ & $3.37\times10^{26}$ & $3.17\times10^{28}$ & $5.29\times10^{19}$ \\ 
$(0,5)\rightarrow(0,3)$ & 9.6894 & $3.54\times10^{3}$ & $1.12\times10^{34}$ & $2.78\times10^{34}$ & $2.90\times10^{34}$ & $2.90\times10^{34}$ & $7.99\times10^{25}$ \\ 
$(1,5)\rightarrow(1,3)$ & 10.2273 & $2.73\times10^{2}$ & $1.35\times10^{30}$ & $1.08\times10^{32}$ & $7.61\times10^{32}$ & $8.85\times10^{32}$ & $2.10\times10^{24}$ \\ 
$(2,5)\rightarrow(2,3)$ & 10.8762 & $2.47\times10^{1}$ & $2.86\times10^{26}$ & $2.49\times10^{29}$ & $1.18\times10^{31}$ & $4.56\times10^{31}$ & $9.96\times10^{22}$ \\ 
$(0,4)\rightarrow(0,2)$ & 12.3006 & $1.75\times10^{3}$ & $7.30\times10^{33}$ & $1.58\times10^{34}$ & $1.55\times10^{34}$ & $1.53\times10^{34}$ & $5.35\times10^{25}$ \\ 
$(1,4)\rightarrow(1,2)$ & 12.9795 & $1.32\times10^{2}$ & $7.97\times10^{29}$ & $4.57\times10^{31}$ & $2.93\times10^{32}$ & $3.39\times10^{32}$ & $1.02\times10^{24}$ \\ 
$(2,4)\rightarrow(2,2)$ & 13.7974 & $1.19\times10^{1}$ & $1.56\times10^{26}$ & $9.89\times10^{28}$ & $3.75\times10^{30}$ & $1.46\times10^{31}$ & $3.69\times10^{22}$ \\ 
$(0,3)\rightarrow(0,1)$ & 17.0595 & $5.63\times10^{2}$ & $2.08\times10^{33}$ & $4.73\times10^{33}$ & $4.58\times10^{33}$ & $4.47\times10^{33}$ & $2.17\times10^{25}$ \\ 
$(1,3)\rightarrow(1,1)$ & 17.9969 & $4.22\times10^{1}$ & $2.03\times10^{29}$ & $8.87\times10^{30}$ & $5.23\times10^{31}$ & $6.75\times10^{31}$ & $2.44\times10^{23}$ \\ 
$(2,3)\rightarrow(2,1)$ & 19.1253 & $3.73\times10^{0}$ & $3.72\times10^{25}$ & $1.82\times10^{28}$ & $5.73\times10^{29}$ & $2.49\times10^{30}$ & $8.72\times10^{21}$ \\ 
$(0,2)\rightarrow(0,0)$ & 28.2559 & $7.90\times10^{1}$ & $1.27\times10^{32}$ & $3.51\times10^{32}$ & $3.63\times10^{32}$ & $3.36\times10^{32}$ & $2.20\times10^{24}$ \\ 
$(1,2)\rightarrow(1,0)$ & 29.8042 & $5.88\times10^{0}$ & $1.12\times10^{28}$ & $4.01\times10^{29}$ & $2.10\times10^{30}$ & $3.35\times10^{30}$ & $1.83\times10^{22}$ \\ 
$(2,2)\rightarrow(2,0)$ & 31.6666 & $0.52\times10^{0}$ &
$1.96\times10^{24}$ & $7.89\times10^{26}$ & $2.16\times10^{28}$ &
$9.82\times10^{28}$ & $5.69\times10^{20}$ \\
\hline 
sum of lines        & --- & --- & $5.32\times10^{34}$ & $4.32\times10^{35}$ &
$8.92\times10^{35}$ & $1.06\times10^{36}$ &  --- \\  
continuum               & --- & $1.00\times10^{0}$ & --- & $8.39\times10^{27}$ &
$1.29\times10^{33}$ & $7.50\times10^{35}$ &  $\sim5\times10^{21}$ \\
total               & --- & --- & $5.32\times10^{34}$ & $4.32\times10^{35}$ &
$8.93\times10^{35}$ & $1.81\times10^{36}$ &  --- \\  
\hline
\end{tabular}
\end{table*}

\end{document}